\documentclass[11pt,a4paper]{article}
\usepackage[english]{babel}
\usepackage{amsmath,amsthm,amssymb,epsfig,latexsym}
\usepackage[numbers,square,sort&compress]{natbib}
\newcommand{\bla}{\bar\lambda}
\newcommand{\bmu}{\bar\mu}
\newcommand{\bu}{\bar u}
\newcommand{\bv}{\bar v}

\newcommand{\ub}[1]{u^{\bs{#1}}}
\newcommand{\bvb}[1]{\bar v^{\bs{#1}}}
\newcommand{\bub}[1]{\bar u^{\bs{#1}}}

\newcommand{\vk}{\varkappa}
\newcommand{\Bom}{S^{\mathbf{\Psi}}}
\newcommand{\Bomj}{S^{\mathbf{J}}}

\newcommand{\GFF}{\mathcal{F}^{\mathbf{\Psi}}}
\newcommand{\GFD}{\mathcal{F}^{\mathbf{J}}}
\newcommand{\HGF}{\;{}_2\hspace{-1pt}F_1}

\newcommand{\bs}[1]{\boldsymbol{#1}}
\newcommand{\so}{\scriptscriptstyle \rm I}
\newcommand{\st}{\scriptscriptstyle \rm I\hspace{-1pt}I}
\newcommand{\sth}{\scriptscriptstyle \rm I\hspace{-1pt}I\hspace{-1pt}I}

\newcommand{\be}[1]{\begin{equation}\label{#1}}
\newcommand{\ba}[1]{\begin{multline}\label{#1}}
\newcommand{\ee}{\end{equation}}
\newcommand{\ea}{\end{eqnarray}}

\newcommand{\dd}{\mathrm{d}}

\newcommand{\tr}{\mathop{\rm tr}}

\newtheorem{prop}{Proposition}[section]
\newtheorem{lemma}{Lemma}[section]
\newtheorem{cor}{Corollary}[section]

\newtheorem{remark}{Remark}[section]
\def\qed{\hfill\nobreak\hbox{$\square$}\par\medbreak}
 \makeatletter
 \@addtoreset{equation}{section}
 \makeatother
 
\begin{document}

\vspace{24pt}

\begin{center}
\begin{LARGE}
{\bf Algebraic Bethe ansatz approach to the correlation functions of the one-dimensional bosons with attraction}
\end{LARGE}

\vspace{50pt}

{\large N.~A.~Slavnov\footnote{nslavnov@mi-ras.ru}}\\
Steklov Mathematical Institute of Russian Academy of Sciences, Moscow, Russia\\

\vspace{80pt}

\centerline{\bf Abstract} \vspace{1cm}
\parbox{12cm}{We consider a model of a one-dimensional Bose gas with attraction. We study ground state equal-time correlation functions
in this model using the algebraic Bethe ansatz. In cases of strong interaction or/and large-volume systems, we obtain very simple explicit formulas
for correlations. }
\end{center}

\section{Introduction\label{S-I}}

The calculation of correlation functions is a central issue in the theory of quantum integrable models \cite{Bax82L,Gau83L,LieM66L}. This problem is of great importance both from a theoretical point of view and for applications to various interesting physical situations. In  several cases, such as free fermions \cite{Ons44,LieSM61,McC68,WuMTB76,McCTW77a} or conformal field theory \cite{BelPZ84}, the correlation functions are quite well studied. However, for many quantum integrable systems, this issue is still far from being completely resolved, despite significant progress achieved especially in recent years.

One of the powerful methods for studying correlation functions of integrable systems is the algebraic Bethe ansatz (ABA) \cite{FadST79,FT79,FadLH96,BogIK93L,Sla22L}. This method was used to calculate correlation functions immediately after its creation \cite{IzeK84,Kor87}. Explicit solutions to the quantum inverse problem \cite{KitMT99,GohK00,MaiT00} further increased the capabilities of this method. Within the framework of this approach, a number of  studies have obtained interesting results for correlation functions in the $XXZ$ Heisenberg chain and a one-dimensional Bose gas with delta interaction \cite{KitMT00,KitMST02a,GohKS04,KitMST05,KitMST05b,KitKMST07,GohBKS07,KitKMST09}. One of the advantages of the ABA is the possibility to obtain compact determinant representations for form factors of local operators \cite{S89,KojKS97,KitMT99}. These representations, in turn, allow one to study correlation functions both analytically \cite{KitKMST11,KitKMST12,KitKMT14} and numerically \cite{CauHM05,PerSCHMWA06,PerSCHMWA07,CauCS07}.

In this paper, we apply the ABA to the study of correlation functions of a one-dimensional Bose gas (the Lieb--Liniger model) with attractive interaction.
The Hamiltonian of this model is given by
\be{Ham}
H=\int_{0}^{L}\Big(\partial_x\Psi^\dagger(x)\partial_x\Psi(x) +c\Psi^\dagger(x)\Psi^\dagger(x)\Psi(x)\Psi(x)\Big)\,\dd x.
\ee
Here $\Psi^\dagger(x)$ and $\Psi(x)$ are canonical Bose fields
\be{CR}
[\Psi(x),\Psi^\dagger(y)]=\delta(x-y),
\ee
acting in a Fock space
\be{Fock}
\Psi(x)|0\rangle=0,\qquad  \langle0|\Psi^\dagger(x)=0.
\ee
The parameter $c$ in \eqref{Ham} is a coupling constant. The case of attraction corresponds to $c<0$. Therefore, in what follows we set $c=-\vk$ and consider $\vk>0$.

The spectral problem for this model was solved in \cite{LiebL63,Lieb63}.
The ground state energy and the wave function in the attractive case were studied in \cite{McG64,CalDG75}.
A formulation of the model in the ABA language was given in \cite{Kul81,IzeKS83}.
A series of multiple integrals for correlation functions of this model was derived in \cite{Kor84,IzeKR87}. In \cite{KojKS97}, a representation for the correlation function was obtained in the form of a Fredholm determinant, depending on auxiliary quantum fields.
Numerical analysis of correlation functions in the Lieb--Liniger model with repulsion was carried out in \cite{CauCS07} using explicit formulas for form factors obtained within the framework of the ABA.

In \cite{CalC07a,CalC07b}, the dynamical correlation functions of one-dimensional bosons with attraction at zero temperature were studied. The calculation method was based on the expansion of correlation functions over form factors, which in turn were previously calculated using the ABA. It was shown that taking into account only two types of excited states already provides good accuracy for correlations.

In this paper, we consider equal-time correlation functions of the one-dimensional bosons with attraction at zero temperature. Unlike dynamical correlations, there is no need to use the form factor expansion in this case. In particular, in \cite{KitKMST07} a closed representation was obtained for the generating function of the density correlations in any quantum integrable model with a 6-vertex $R$-matrix\footnote{A similar representation for the field correlation functions is obtained in this work.}. This result is quite applicable to the Lieb--Liniger model. It should be said, however, that the above-mentioned representation is very cumbersome and contains summation over all possible partitions of the Bethe parameters. In \cite{KitKMST07}, this sum over partitions was reduced to a multiple contour integral of Cauchy type (master-equation \cite{KitMST05}), which in turn generates numerous integral representations for correlation functions in the thermodynamic limit \cite{KitKMST07,KitKMST09}.

However, in the case of one-dimensional bosons with attraction, there is a significant simplification of all formulas in the $\vk L\gg 1$ regime. It is based on a very simple ground state structure. The Bethe equations, which characterize the spectrum of the Hamiltonian, can be solved explicitly up to exponentially small corrections at $\vk L\gg 1$. At the same time, just a few numbers of terms survive in the representations for correlation functions. Moreover, in the case of an arbitrary solution of the Bethe equations, these terms contain rather complex determinants. However, for the ground state of one-dimensional bosons with attraction, these determinants are also simplified and can be calculated explicitly. It was this feature that the authors of \cite{CalC07a,CalC07b} took advantage of, obtaining explicit and very simple formulas for form factors. In our case, all the simplifications listed above lead to the fact that we obtain explicit results for the correlation functions of densities and fields up to exponentially small corrections at $\vk L\gg 1$.

We present now the main results of the paper. The ground state correlation function of fields is given by
\be{CF-Fie-res1}
\langle \Psi^\dagger(x)\Psi(0)\rangle=\frac1L e^{-\vk x(N-1)/2}
\sum_{\ell=1}^N  e^{-\vk x(\ell-1)(N-\ell)}
\left(1+\frac{\vk x}2(N-2\ell+1)^2\right)+O\big((\vk L)^{-\infty}\big),
\ee
where $x>0$, and $N$ is the number of particles in the ground state. The density correlation function has the following form:
\be{veryfin-res}
\langle j(x)j(0)\rangle=
\frac{1}{\vk L}\frac{\partial^2}{\partial x^2}\sum_{\ell=1}^{N-1} e^{-\vk x\ell(N-\ell)}
\left(1+\frac{x\vk}2(N-2\ell)^2\right)+O\big((\vk L)^{-\infty}\big),
\ee
where the density operator $j(x)$ is
\be{dens}
j(x)=\Psi^\dagger(x)\Psi(x).
\ee

The paper is organized as follows. In section~\ref{S-ABA}, we recall the basic concepts of the ABA, introduce the notation, define Bethe vectors, and introduce the composite model. We also describe there the ground state of the attractive one-dimensional bosons model. In section~\ref{S-SPBV}, we give determinant representations for scalar products of Bethe vectors and some form factors.
Section~\ref{S-RCF} contains representations for generating functions of the density and the field correlations. These representations are key to our approach. In section~\ref{S-SCFG}, we consider a special case of the determinants introduced above in section~\ref{S-SPBV}. Finally, in section~\ref{S-GSCF}, we calculate the ground state correlation functions of densities and fields. Appendix~\ref{A-RGF} contains a derivation of the representation for the generating function of the field correlations.

\section{Algebraic Bethe ansatz\label{S-ABA}}

In this section, we briefly recall the basic information about the ABA, which is necessary for further understanding of the paper. The reader can find a more detailed presentation of this method in the original works \cite{FadST79,FT79} or in numerous monographs (see, for example, \cite{BogIK93L,Sla22L}). We mainly adhere to the terminology and notation of \cite{Sla22L}.

One of the main objects of the ABA is an $R$-matrix $R(u,v)$. The one-dimensional bosons model is described by
the $GL(2)$-invariant $R$-matrix acting in the tensor product of two auxiliary spaces $V_1\otimes V_2$, where
$V_k\sim\mathbb{C}^2$, $k=1,2$:
 \be{R-mat}
 R(u,v)=\mathbf{I}+g(u,v)\mathbf{P},\qquad g(u,v)=\frac{\eta}{u-v}.
 \ee
In the above definition, $\mathbf{I}$ is the identity matrix in $V_1\otimes V_2$, $\mathbf{P}$ is the permutation matrix
that exchanges $V_1$ and $V_2$, and $\eta$ is a constant. For the model of  one-dimensional bosons, it is related to the coupling constant by $\eta=i\vk$.

Another important object of the ABA is a monodromy matrix $T(u)$:
\be{Monod}
T(u)=\begin{pmatrix} A(u)& B(u)\\
C(u)& D(u)\end{pmatrix}.
\ee
The entries of the monodromy matrix are operators acting in the quantum space $\mathcal{H}$ of the Hamiltonian \eqref{Ham}. They depend on a complex parameter $u$
and satisfy the following algebra:
\be{RTT}
R_{12}(u_1,u_2)T_1(u_1)T_2(u_2)=T_2(u_2)T_1(u_1)R_{12}(u_1,u_2).
\ee
Equation \eqref{RTT} holds in the tensor product $V_1\otimes V_2\otimes\mathcal{H}$,
where $V_k\sim\mathbb{C}^2$, $k=1,2$, are the auxiliary linear spaces. The $R$-matrix acts non-trivially in $V_1\otimes V_2$, the matrices $T_k(u)$ act non-trivially in
$V_k\otimes \mathcal{H}$.

The trace of the monodromy matrix in the auxiliary space $\tr T(u)=A(u)+D(u)$ is called a transfer matrix. It is a generating
functional of integrals of motion of the model. The eigenvectors of the transfer matrix are
called on-shell Bethe vectors (or simply on-shell vectors). They can be parameterized by a set of complex parameters
satisfying Bethe equations (see below).

A necessary condition for the applicability of the ABA is the existence in the space $\mathcal{H}$ of a special vector called pseudovacuum. In the case of the Lieb--Liniger model, this vector coincides with the Fock vacuum $|0\rangle$ \eqref{Fock}. It is an eigenvector to the diagonal matrix elements of the monodromy matrix and is annihilated by the operator $C(u)$:
 \be{Tjj}
 \begin{aligned}
 &A(u)|0\rangle=a(u)|0\rangle,\\
&D(u)|0\rangle=d(u)|0\rangle,\\
 &C(u)|0\rangle=0.
 \end{aligned}
 \ee
In the one-dimensional bosons model
\be{ad}
a(u)=e^{-iLu/2}, \qquad d(u)=e^{iLu/2}.
\ee
Below we will often use the ratio of the functions $a(u)$ and $d(u)$
\be{r-ad}
r(u)=\frac{a(u)}{d(u)}=e^{-iLu}.
\ee

We will also need a dual pseudovacuum vector $\langle 0|$ belonging to the dual space $\mathcal{H}^*$. It possesses similar properties
 \be{Tjjd}
 \begin{aligned}
 &\langle0|A(u)=a(u)\langle0|,\\
 &\langle0|D(u)=d(u)\langle0|,\\
 &\langle0|B(u)=0,
 \end{aligned}
 \ee
with the same functions $a(u)$ and $d(u)$ \eqref{ad}.

\subsection{Notation\label{SS-N}}

Before moving on we say some words about the notation used in the paper. Besides the function $g(u,v)$ \eqref{R-mat} we also introduce  rational functions
\be{fg}
 f(u,v)=1+g(u,v)=\frac{u-v+\eta}{u-v}
\ee
and
\be{univ-not}
  h(u,v)=\frac{f(u,v)}{g(u,v)}=\frac{u-v+c}{\eta},\qquad  t(u,v)=\frac{g(u,v)}{h(u,v)}=\frac{\eta^2}{(u-v)(u-v+c)}.
\ee

We will permanently deal with sets of parameters. These sets are always denoted  by bars: $\bu$, $\bla$ etc.
Individual elements of the sets are denoted by subscripts: $u_j$, $\lambda_k$ etc. As a rule, the number of elements in the
sets is not shown explicitly in the equations; however, we give these cardinalities in
special comments after the formulas.
We also use a special notation  for subsets with one element omitted  $\bu_k=\bu\setminus u_k$, $\bla_j=\bla\setminus\lambda_j$,
and so on.

To avoid  formulas being too cumbersome we use shorthand notation for products of scalar
functions or commuting operators. Namely, if functions $a$, $d$, $r$, or  the monodromy matrix entries depend
on sets of parameters, this means that one should take the product over the corresponding set.
For example,
 \be{SH-prod0}
 d(\bar u)=\prod_{u_j\in\bar u} d(u_j);\quad
 r(\bar v_j)= \prod_{\substack{v_k\in\bar v\\v_k\ne v_j}} r(v_k);\quad
 B(\bla)=\prod_{\lambda_j\in\bla}B(\lambda_j).
 \ee
The last equation is well defined since $[B(u),B(v)]=0$ due to \eqref{RTT}. We will extend this  convention on the shorthand notation for the products to some other functions and operators that will appear as we proceed.

We use the same convention for the rational functions introduced above
 \be{SH-prod}
 g(u_k, \bar v)= \prod_{v_j\in\bar v} g(u_k, v_j);\quad
 h(u,\bv_j)=\prod_{\substack{v_k\in\bar v\\v_k\ne v_j}} h(u,v_k);\quad
 f(\bla,\bmu)=\prod_{\lambda_j\in\bla}\prod_{\mu_k\in\bmu} f(\lambda_j,\mu_k).
 \ee
By definition, any product over the empty set is equal to $1$. A double product is equal to $1$ if at least one of the sets
is empty.

Finally, let us introduce a new notation for special products of the functions $g(u,v)$
\begin{equation}\label{07-SP-defD}
\Delta(\bu)=\prod_{1\le j<k\le N}g(u_k,u_j),\qquad
\Delta'(\bu)=\prod_{1\le j<k\le N}g(u_j,u_k),
\end{equation}
where $N=\#\bu\ge 2$. It is clear that $\Delta'(\bu)=(-1)^{N(N-1)/2}\Delta(\bu)$. For $N=0,1$, we set $\Delta(\bu)=\Delta'(\bu)=1$ by definition.

\subsection{Bethe vectors\label{SS-BVSP}}

Eigenvectors of the transfer matrix $A(z)+D(z)$ can be obtained by the successive action of the $B$-operators on the pseudovacuum.  Let $\bu=\{u_1,\dots,u_N\}$, where
$N=0,1,\dots$. Then a vector $B(\bu)|0\rangle$ is an eigenvector of the transfer matrix
\be{EVP}
\big(A(z)+D(z)\big)B(\bu)|0\rangle=\tau(z;\bu)B(\bu)|0\rangle,
\ee
provided the parameters $\bu$ satisfy a system of Bethe equations
\be{BE}
 r(u_j)=\frac{f(u_j,\bu_j)}{f(\bu_j,u_j)}, \qquad j=1,\dots,N.
\ee
Then
\be{EV}
\tau(z;\bu)=a(z)f(\bu,z)+d(z)f(z,\bu).
\ee
We call vectors $B(\bu)|0\rangle$ on-shell Bethe vectors if the set $\bu$ satisfies equations \eqref{BE}. Otherwise, if $\bu$ consists of arbitrary complex numbers,
we call the corresponding vectors $B(\bu)|0\rangle$ off-shell Bethe vectors.

One can also define dual Bethe vectors belonging to the dual space $\mathcal{H}^*$. They are constructed by the right successive action of the $C$-operators
on the dual pseudovacuum $\langle0|$. Then
\be{D-EVP}
\langle0|C(\bu)\big(A(z)+D(z)\big)=\tau(z;\bu)\langle0|C(\bu),
\ee
provided $\bu$ satisfy Bethe equations \eqref{BE}. In this case, we call $\langle0|C(\bu)$ a dual on-shell Bethe vector. If $\bu$ consists of arbitrary complex numbers,
then $\langle0|C(\bu)$ is called a dual off-shell Bethe vector.

We also will consider twisted Bethe equations
\be{tBE}
 r(v_j)=e^\beta\frac{f(v_j,\bv_j)}{f(\bv_j,v_j)}, \qquad j=1,\dots,N,
\ee
where $\beta$ is a complex number. The corresponding vectors $\langle0|C(\bv)$ and $B(\bv)|0\rangle$ then are eigenvectors of the twisted transfer matrix
$A(z)+e^\beta D(z)$. We call them twisted (dual) on-shell Bethe vectors.

Below, it will be convenient for us to normalize the elements of the monodromy matrix to the function $d(u)$ \eqref{Tjj}, \eqref{ad}
\be{bold}
\bs{B}(u)=\frac{B(u)}{d(u)}, \qquad \bs{C}(u)=\frac{C(u)}{d(u)},
\ee
and construct Bethe vectors using
the $\bs{B}$- and $\bs{C}$-operators
\be{bolBV}
\bs{B}(\bu)|0\rangle, \qquad\qquad \langle0|\bs{C}(\bu).
\ee
Here we have extended the convention on the shorthand notation \eqref{SH-prod0} to the products of operators $\bs{B}(u)$ and $\bs{C}(u)$.
With this normalization, scalar products and correlation functions automatically depend on the function $r(u)$ \eqref{r-ad}.

\subsection{Composite model\label{SS-CM}}

Actions of the Bose fields $\Psi(0)$ and $\Psi^\dagger(0)$ on off-shell Bethe vectors were calculated in \cite{IzeKR87,KojKS97}
\be{Psiact}
\begin{aligned}
&\Psi(0)\bs{B}(\bu)|0\rangle=\sqrt{\vk}\sum_{j=1}^Nr(u_j)f(\bu_j,u_j)\bs{B}(\bu_j)|0\rangle,\\
&\langle0|\bs{C}(\bu)\Psi^\dagger(0)=-\sqrt{\vk}\sum_{j=1}^N f(u_j,\bu_j)\langle0|\bs{C}(\bu_j).
\end{aligned}
\ee
Here $\bu=\{u_1,\dots,u_N\}$ are arbitrary complex numbers.

\begin{remark}\label{rem-a0}
The action of the fields $\Psi(0)$ and $\Psi^\dagger(0)$ on Bethe vectors can be obtained using the model of one-dimensional bosons on a lattice \cite{IzeK81a}. In this model, there is some ambiguity in the definition of the $B$ and $C$-operators. This ambiguity is partially preserved in the continuous limit. In particular, we can replace $B\to\omega B$ and $C\to\omega^{-1}C$, where $\omega$ is an arbitrary non-zero number. The transformed monodromy matrix still satisfies the $RTT$-relation \eqref{RTT} and has the same pseudovacuum vector as the original one. Obviously, additional factors appear in formulas \eqref{Psiact} under this transformation. However, they compensate each other in quadratic expressions. We choose a normalization of the $B$ and $C$-operators such that $C(u)=-B^\dagger(u^*)$.
\end{remark}

To find the actions of $\Psi(x)$ and $\Psi^\dagger(x)$ on off-shell Bethe vectors, we
introduce a composite model \cite{IzeK84}. In this model, we present the total monodromy matrix $T(u)$ as a product of two partial monodromy matrices
\be{T-TT}
T(u)=T^{(2)}(u)T^{(1)}(u).
\ee
Here the total monodromy matrix $T(u)$ is related to the interval $[0,L]$, while the partial monodromy matrices $T^{(1)}(u)$ and $T^{(2)}(u)$ respectively  are associated with the intervals $[0,x[$ and $[x,L]$. Both $T^{(1)}(u)$ and $T^{(2)}(u)$ have the form \eqref{Monod}
 \begin{equation}\label{11-ABAT-2s}
 T^{(j)}(u)=\left(
\begin{array}{cc}
A^{(j)}(u)&B^{(j)}(u)\\
C^{(j)}(u)&D^{(j)}(u)
\end{array}\right), \qquad j=1,2.
\end{equation}
Each of the partial monodromy matrices satisfies the  $RTT$-relation \eqref{RTT}
with the $R$-matrix \eqref{R-mat}.  The matrix elements of the different partial matrices
commute with each other. Each of the matrices
$T^{(j)}(u)$ possesses a vector $|0\rangle^{(j)}$ and a dual vector
$\langle0|^{(j)}$, where $|0\rangle=|0\rangle^{(2)}\otimes|0\rangle^{(1)}$.
The action formulas of the operators on these vectors are similar to the action formulas for the total monodromy matrix
 \begin{equation}\label{11-action-2s}
 \begin{aligned}
& A^{(j)}(u)|0\rangle^{(j)}=a^{(j)}(u)|0\rangle^{(j)},&\qquad &\langle0|^{(j)}A^{(j)}(u)=a^{(j)}(u)\langle0|^{(j)},\\
&  D^{(j)}(u)|0\rangle^{(j)}=d^{(j)}(u)|0\rangle^{(j)},&\qquad & \langle0|^{(j)}D^{(j)}(u)=d^{(j)}(u)\langle0|^{(j)},\\
&   C^{(j)}(u)|0\rangle^{(j)}=0,&\qquad& \langle0|^{(j)}B^{(j)}(u)=0,
\end{aligned}
\end{equation}
where functions $a^{(j)}(u)$ and $d^{(j)}(u)$ in the Lieb--Liniger model are
\be{ad-x}
\begin{aligned}
&a^{(1)}(u)=e^{-ixu/2},& \qquad & d^{(1)}(u)=e^{ixu/2},\\
&a^{(2)}(u)=e^{-i(L-x)u/2},& \qquad & d^{(2)}(u)=e^{i(L-x)u/2}.
\end{aligned}
\ee
We will also often use the ratio of the functions $a^{(j)}(u)$ and $d^{(j)}(u)$
\be{rr}
r^{(1)}(u)=e^{-iux}, \qquad r^{(2)}(u)=e^{-iu(L-x)}=\frac{r(u)}{r^{(1)}(u)}.
\ee

In complete analogy with \eqref{bold}, we introduce
\begin{equation}\label{13-boldBC}
\bs{B}^{(j)}(u)=\frac{B^{(j)}(u)}{d^{(j)}(u)},\qquad \bs{C}^{(j)}(u)=\frac{C^{(j)}(u)}{d^{(j)}(u)},
\qquad j=1,2,
\end{equation}
and extend the convention on the shorthand notation to the products of partial operators and partial vacuum eigenvalues. For example,
\begin{equation}\label{04-Conv}
\bs{B}^{(j)}(\bar{v})=\prod_{v_k\in\bar{v}}\bs{B}^{(j)}(v_k),\qquad r^{(j)}(\bar{u})=\prod_{u_l\in\bar{u}} r^{(j)}(u_l),
\end{equation}
and so on.

The key formulas of the composite model are expressions of the total off-shell Bethe vectors $\bs{B}(\bu)|0\rangle$ and $\langle0|\bs{C}(\bu)$ in
terms of partial ones $\bs{B}^{(j)}(\bu)|0\rangle$ and $\langle0|\bs{C}^{(j)}(\bu)$. These formulas were obtained in \cite{IzeK84}
(see also \cite{BogIK93L,Sla22L}):
\begin{equation}\label{13-IK}
\bs{B}(\bu)|0\rangle=\sum_{\bu\mapsto\{\bu_{\so},\bu_{\st}\}}
   r^{(2)}(\bu_{\so})\bs{B}^{(1)}(\bu_{\so})|0\rangle^{(1)}\bs{B}^{(2)}(\bu_{\st})|0\rangle^{(2)}
 f(\bu_{\st},\bu_{\so}),
 \end{equation}
and
\begin{equation}\label{13-IK-d}
 \langle0| \bs{C}(\bu)=\sum_{\bu\mapsto\{\bu_{\so},\bu_{\st}\}}
r^{(1)}(\bu_{\st})
\langle0|^{(1)}\bs{C}^{(1)}(\bu_{\so})\langle0|^{(2)}\bs{C}^{(2)}(\bu_{\st})
\;  f(\bu_{\so},\bu_{\st}).
 \end{equation}
Here the set $\bu$ consists of arbitrary complex numbers. The sums in \eqref{13-IK} and \eqref{13-IK-d} are taken over all possible partitions  of the set $\bu$ into two disjoint subsets $\bu_{\so}$ and $\bu_{\st}$ (including the partition into the empty set and the full set).

It is easy to see that the intermediate point $x$ plays the same role for the matrix $T^{(2)}(u)$ as point $0$ does for the matrix $T(u)$. Therefore, the action
of the fields $\Psi(x)$ and $\Psi^\dagger(x)$  on vectors $\bs{B}^{(2)}(\bu)|0\rangle$ and $\langle0|\bs{C}^{(2)}(\bu)$ is similar to actions \eqref{Psiact}
\be{Psiact2}
\begin{aligned}
&\Psi(x)\bs{B}^{(2)}(\bu)|0\rangle=\sqrt{\vk}\sum_{j=1}^Nr^{(2)}(u_j)f(\bu_j,u_j)\bs{B}^{(2)}(\bu_j)|0\rangle,\\
&\langle0|\bs{C}^{(2)}(\bu)\Psi^\dagger(x)=-\sqrt{\vk}\sum_{j=1}^N f(u_j,\bu_j)\langle0|\bs{C}^{(2)}(\bu_j).
\end{aligned}
\ee
Equations \eqref{Psiact2} together with formulas \eqref{13-IK} and \eqref{13-IK-d} provide the actions of $\Psi(x)$ and $\Psi^\dagger(x)$  on off-shell Bethe vectors.

\subsection{Ground state\label{SS-GS}}

In the model of one-dimensional bosons with repulsion, all solutions of the Bethe equations are real. In the case of attraction, in addition to real roots, complex conjugate pairs can also arise. They form so-called string solutions.

The Bethe equations have the following form:
\be{BE-NS}
 e^{iLu_j}=\prod_{\substack{k=1\\ k\ne j}}^N \frac{u_j-u_k-i\vk}{u_j-u_k+i\vk}, \qquad j=1,\dots,N.
\ee
Twisted Bethe equations have the form
\be{tBE-NS}
 e^{iLv_j}=e^{-\beta}\prod_{\substack{k=1\\ k\ne j}}^N \frac{v_j-v_k-i\vk}{v_j-v_k+i\vk}, \qquad j=1,\dots,N.
\ee
It is easy to see that if a set $\bu$ is a solution to the ordinary Bethe equations \eqref{BE-NS}, then the set $\bv$, where $v_j=u_j+i\beta/L$, solves the twisted Bethe equations \eqref{tBE-NS}.

If $\bu$ is a solution to the Bethe equations, then the eigenvalue of the Hamiltonian on the on-shell vector $\bs{B}(\bu)|0\rangle$ has the form
\be{EiVal}
E[\bu]=\sum_{j=1}^N u_j^2.
\ee
In this paper, we consider correlation functions in the ground state. We will denote the corresponding roots of the Bethe equations by $\bla=\{\lambda_1,\dots,\lambda_N\}$,
where $N=0,1,\dots$. This solution forms one string centered at zero. In other words, all $\lambda_j$ are purely imaginary so that $\lambda_j=-\lambda_{N-j+1}$.

Let $\lambda_j=i\vk\tilde \lambda_j$, where $\tilde \lambda_j$ are real numbers. Then Bethe equations for the set $\{\tilde \lambda_1,\dots,\tilde \lambda_N\}$ take the form
\be{BE-NS-til}
 e^{-\vk L\tilde \lambda_j}=\prod_{\substack{k=1\\ k\ne j}}^N \frac{\tilde \lambda_j-\tilde \lambda_k-1}{\tilde \lambda_j-\tilde \lambda_k+1}, \qquad j=1,\dots,N.
\ee
We see that in the $\vk L\gg 1$ regime, the lhs of these equations exponentially go to either zero or infinity\footnote{Except for the root $\tilde\lambda_{(N+1)/2}=0$, which occurs for $N$ odd.}. A simple analysis shows \cite{McG64,Tha81,Tak99} that in this case, the interval between two adjacent roots $\tilde \lambda_{j+1}$ and $\tilde \lambda_j$ should go to $1$. Thus,
\be{spec-sol0}
\tilde \lambda_j=\tilde \lambda_1+j-1+\epsilon_j,\qquad  \epsilon_j=O\left((\vk L)^{-\infty}\right), \qquad j=1,\dots,N,
\ee
where $\tilde \lambda_1=(1-N)/2$. The corrections $\epsilon_j$ have the form $\epsilon_j\approx e^{-\vk L a_j}$, where $a_j>0$. In all our calculations below we neglect these
exponentially small corrections.

We will also deal with a twisted ground state corresponding to the solution of the twisted Bethe equations. We denote the corresponding roots by $\bmu=\{\mu_1,\dots,\mu_N\}$. They have the form $\mu_j=\lambda_j+i\beta/L$,
where $\lambda_j$ are the roots of the ground state.

To conclude this section, we recall that $C(u)=-B^\dagger(u^*)$ in the attractive Lieb--Liniger model. Therefore, we have for the ground state solution
\be{Hermit}
\langle0|\bs{C}(\bla)=(-1)^N \big(\bs{B}(\bla)|0\rangle\big)^\dagger.
\ee
In the ABA method, it is customary to call the value $\langle0|\bs{C}(\bu)\bs{B}(\bu)|0\rangle$ the squared norm of the Bethe vector. We see that the scalar product
$\langle0|\bs{C}(\bla)\bs{B}(\bla)|0\rangle$ may differ in sign from the true squared norm. However, this difference is insignificant, since the $(-1)^N$ factor cancels in the expressions for the correlation functions (see \eqref{CF-Den}, \eqref{CF-Fie}).

\section{Scalar products of Bethe vectors and form factors\label{S-SPBV}}

In this section, we present some formulas for scalar products of Bethe vectors. We also give formulas for form factors that are a direct consequence of the determinant representations for the scalar products.

\subsection{General scalar products\label{SS-GSP}}

We call the scalar product the pseudovacuum average
\begin{equation}\label{SP-0def}
\bs{S}_N(\bv,\bu)= \langle0|\bs{C}(\bv) \bs{B}(\bu)|0\rangle.
\end{equation}
We do not impose any constraint
on the sets $\bv$ and $\bu$. Thus, generically we deal with off-shell Bethe vectors in \eqref{SP-0def}.

To describe the scalar product \eqref{SP-0def} we first introduce the notion of the highest coefficient \cite{Kor82}. Namely, for two sets of variables of equal cardinality
$\bv=\{v_1,\dots,v_N\}$ and $\bu=\{u_1,\dots,u_N\}$, the highest coefficient $\mathbb{K}_N(\bv|\bu)$ is  \cite{Ize87}
\be{HC}
\mathbb{K}_N(\bv|\bu)=\Delta'(\bv)\Delta(\bu)h(\bv,\bu)\det_N t(v_j,u_k).
\ee
Then the scalar product \eqref{SP-0def} is given by \cite{Kor82}
\begin{equation}\label{SP1}
 \bs{S}_{N}(\bv,\bu)=
 \sum_{\substack{\bv\mapsto\{\bv_{\so},\bv_{\st}\}\\\bu\mapsto\{\bu_{\so},\bu_{\st}\}}}
r(\bv_{\st})r(\bu_{\so})
\mathbb{K}_{N_{\so}}( \bv_{\so}|\bu_{\so})\mathbb{K}_{N_{\st}}( \bu_{\st}|\bv_{\st})\,f(\bu_{\st},\bu_{\so})f(\bv_{\so},\bv_{\st}).
\end{equation}
The sum in \eqref{SP1} is taken over all possible partitions  of the set $\bu$ into two disjoint subsets $\bu_{\so}$ and $\bu_{\st}$ and the set $\bv$ into two disjoint subsets $\bv_{\so}$ and $\bv_{\st}$, so that $\#\bu_{\so}=\#\bv_{\so}=N_{\so}$, $\#\bu_{\st}=\#\bv_{\st}=N_{\st}$, $N_{\so}+N_{\st}=N$, and $N_{\so}=0,\dots,N$.
\begin{remark}\label{rem-a01}
Restrictions on the cardinalities of the subsets arise because the highest coefficients exist only for sets of the same cardinality.
\end{remark}
Equation \eqref{SP1} describes the scalar product of two arbitrary off-shell Bethe vectors. However, if at least one of the vectors is on-shell, the sum over the partitions turns into a single determinant of an $N\times N$ matrix \cite{S89}. In what follows we will need some particular cases of the scalar products.

\subsection{Density form factor\label{SS-DFF}}

Let $\#\bu=\#\bv=N$. We introduce
\be{Fn-def}
\Bomj_N(\bv,\bu|\beta)=\Delta'(\bv)\Delta(\bu)h(\bu,\bv)\det_N \mathcal{M}^{\mathbf{J}}_{jk}(\bv,\bu),
\ee
where
\be{MJjk}
\mathcal{M}^{\mathbf{J}}_{jk}(\bv,\bu)=e^\beta t(u_k,v_j)+t(v_j,u_k)\frac{h(u_k,\bu)h(\bv,u_k)}{h(\bu,u_k)h(u_k,\bv)}.
\ee
Function $\Bomj_N(\bv,\bu|\beta)$ describes the scalar product of a twisted dual on-shell Bethe vector $\langle0|\bs{C}(\bv)$ (that is, the set $\bv$
satisfies equations \eqref{tBE}) and ordinary on-shell Bethe vector $\bs{B}(\bu)|0\rangle$ (that is, the set $\bu$
satisfies equations \eqref{BE}). In particular, it gives an expression for the norm of an on-shell Bethe vector in the limit $\beta\to0$.
The function \eqref{Fn-def} also has a property $\Bomj_N(\bv,\bu|\beta)=e^{\beta N}\Bomj_N(\bu,\bv|-\beta)$, which follows from representation \eqref{SP1} for the scalar product.

Representation \eqref{Fn-def} is valid for any integrable model with the
$R$-matrix \eqref{R-mat}. In the case of the Lieb--Liniger model, this scalar product is closely related to the form factor
of the density operator \eqref{dens} \cite{S89}
\begin{equation}\label{ForfacJ-res}
\langle0|\bs{C}(\bv)\;j(x)\; \bs{B}(\bu)|0\rangle
=\frac{\partial}{\partial x}\left(\prod_{j=1}^Ne^{ix(u_j-v_j)}\right)\frac{\partial}{\partial\beta}\Bomj_N(\bv,\bu|\beta)\Bigr|_{\beta=0}.
\end{equation}
Here $\langle0|\bs{C}(\bv)$ and $\bs{B}(\bu)|0\rangle$ are two on-shell Bethe vectors. We will see that the two-point function of densities
\eqref{veryfin-res} also is expressed in terms of the function $\Bomj_N(\bv,\bu|\beta)$.

Recall some properties of $\Bomj_N(\bv,\bu|\beta)$ for generic $\bv$ and $\bu$ \cite{S89} (see also \cite{Sla22L}). It is a rational function of $\bv$ and $\bu$, which is symmetric separately over $\bv$ and $\bu$. It vanishes if
one of its arguments goes to infinity. It has poles at $v_j=u_k$, $j,k=1,\dots,N$, and has no other poles. In particular, it is not singular at
$h(u_j,u_k)=0$ for $j,k=1,\dots,N$, despite the matrix elements $\mathcal{M}^{\mathbf{J}}_{jk}$ have poles
in this case. The proof of this property can be found in \cite{S89} (see also \cite{Sla22L}). However, since this property plays a very important role, we also provide the proof here.

We present the determinant of $\mathcal{M}^{\mathbf{J}}_{jk}$ as follows:
\be{extract}
\det_{N}\mathcal{M}^{\mathbf{J}}_{jk}(\bv,\bu)=\frac1{h(\bu,\bu)} \det_{N}\Big(h(\bu,u_k)\mathcal{M}^{\mathbf{J}}_{jk}(\bv,\bu)\Big).
\ee
Here
\be{tM-e}
h(\bu,u_k)\mathcal{M}^{\mathbf{J}}_{jk}(\bv,\bu)=e^\beta t(u_k,v_j)h(\bu,u_k)+ t(v_j,u_k)
\frac{h(u_k,\bu)h(\bv,u_k)}{h(u_k,\bv)}.
\ee
Due to the symmetry of $\Bomj_N(\bv,\bu|\beta)$ over $\bu$, it is enough to prove the absence of the pole at $u_2=u_1+\eta$. Then $h(u_1,u_2)=0$, and the prefactor in the rhs of \eqref{extract} becomes singular. On the other hand,
\be{tM-e1}
 h(\bu,u_1)\mathcal{M}^{\mathbf{J}}_{j1}(\bv,\bu)=e^\beta t(u_1,v_j)h(\bu,u_1),
\ee
and
\be{tM-e2}
 h(\bu,u_2)\mathcal{M}^{\mathbf{J}}_{j2}(\bv,\bu)= t(v_j,u_2)
\frac{h(u_2,\bu)h(\bv,u_2)}{h(u_2,\bv)}=t(u_1,v_j)
\frac{h(u_2,\bu)h(\bv,u_2)}{h(u_2,\bv)}.
\ee
Here we used $t(v_j,u_2)=t(u_1,v_j)$ for $u_2=u_1+\eta$. We see that the first and the second columns of the matrix are proportional to each other. Hence,
its determinant vanishes.

Finally, one more important property of the function $\Bomj_N(\bv,\bu|\beta)$ is the following recursion:
\be{Fn-rec}
\Bomj_N(\bv,\bu|\beta)=\sum_{j=1}^N g(v_N,u_j)\Bigl(f(u_j,\bu_j)f(\bv_N,u_j)-e^\beta f(\bu_j,u_j)f(u_j,\bv_N)\Bigr)\Bomj_{N-1}(\bv_N,\bu_j|\beta).
\ee
It is easy to see that this recursion is nothing but the expansion of $\Bomj_N(\bv,\bu|\beta)$ with respect to the poles at $v_N=u_j$ ($j=1,\dots, N$). Equation \eqref{Fn-rec}
allows one to construct $\Bomj_N(\bv,\bu|\beta)$ successively starting from
\be{Fn1-def}
\Bomj_1(v,u|\beta)=g(u,v)\left(e^\beta-1\right).
\ee

\subsection{Field form factor\label{SS-FFF}}

Let $\#\bu=N$ and $\#\bv=N-1$. We introduce
\be{detrepG2-a00}
\Bom_N(\bv,\bu|\beta)=\Delta'(\bv)\Delta(\bu)h(\bu,\bv)
\det_{N}\mathcal{M}^{\mathbf{\Psi}}_{jk}(\bv,\bu),
\ee
where matrix $\mathcal{M}^{\mathbf{\Psi}}_{jk}(\bv,\bu)$ is given by 
\be{tM-a00}
\begin{aligned}
& \mathcal{M}^{\mathbf{\Psi}}_{jk}(\bv,\bu)=e^\beta t(u_k,v_j)- t(v_j,u_k)
\frac{h(u_k,\bu)h(\bv,u_k)}{h(\bu,u_k)h(u_k,\bv)}, \qquad j<N,\\
&\mathcal{M}^{\mathbf{\Psi}}_{jk}(\bv,\bu)= \frac{h(u_k,\bu)}{h(u_k,\bv)},\qquad j=N.
\end{aligned}
\ee
Function $\Bom_N(\bv,\bu|\beta)$ is closely related to the field form factor\footnote{%
Matrix element \eqref{ForfacP-res} contains different numbers of the $B$- and $C$-operators. Therefore, the formula for the form factor depends on the initial normalization of these operators (see Remark~\ref{rem-a0}). } \cite{KojKS97}
\begin{equation}\label{ForfacP-res}
\langle0|\bs{C}(\bv)\;\Psi(x)\; \bs{B}(\bu)|0\rangle
=\sqrt{\vk}\left(\prod_{j=1}^Ne^{ixu_j}\right)\left(\prod_{j=1}^{N-1}e^{-ix v_j}\right)\Bom_N(\bv,\bu|\beta)\Bigr|_{\beta=0}.
\end{equation}
Here $\langle0|\bs{C}(\bv)$ and $\bs{B}(\bu)|0\rangle$ are two on-shell Bethe vectors. The correlation function of fields
\eqref{CF-Fie-res1} also is expressed in terms of the function $\Bom_N(\bv,\bu|\beta)$.

Properties of $\Bom_N(\bv,\bu|\beta)$ for generic $\bv$ and $\bu$ are very similar to the ones of $\Bomj_N(\bv,\bu|\beta)$ (see \cite{Sla22L}). It is a rational function of $\bv$ and $\bu$, which is symmetric separately over $\bv$ and $\bu$. It vanishes if
one of $v_j\to\infty$. It has poles at $v_j=u_k$, $j=1,\dots,N-1$, $k=1,\dots,N$, however it has no  poles at
$h(u_j,u_k)=0$ for $j,k=1,\dots,N$. The proof is completely analogous to the one considered above for the function $\Bomj_N(\bv,\bu|\beta)$.

Expanding $\Bom_N(\bv,\bu|\beta)$ over the poles at $v_{N-1}=u_k$, $k=1,\dots,N$, we obtain the following recursion:
\begin{multline}\label{rec-G1}
\Bom_{N}(\bv,\bu|\beta)=\sum_{j=1}^N  g(u_j,v_{N-1})
\Big(e^\beta f(\bu_j,u_j)f(u_j,\bv_{N-1})\\
-f(u_j,\bu_j)f(\bv_{N-1},u_j)\Big)\Bom_{N-1}(\bv_{N-1},\bu_j|\beta),
\end{multline}
with the initial condition
\be{in-cond}
\Bom_{1}(\emptyset,u|\beta)=1.
\ee

\section{Representations for correlation functions\label{S-RCF}}

In this paper, we study the zero temperature correlation function of densities
\be{CF-Den}
\langle j(x)j(0)\rangle=\frac{\langle0|\bs{C}(\bla)\; j(x)j(0)\;\bs{B}(\bla)|0\rangle}
{\langle0|\bs{C}(\bla)\bs{B}(\bla)|0\rangle},
\ee
and the correlation function of fields
\be{CF-Fie}
\langle \Psi^\dagger(x)\Psi(0)\rangle=\frac{\langle0|\bs{C}(\bla)\; \Psi^\dagger(x)\Psi(0)\;\bs{B}(\bla)|0\rangle}
{\langle0|\bs{C}(\bla)\bs{B}(\bla)|0\rangle}.
\ee
Here the set $\bla=\{\lambda_1,\dots,\lambda_N\}$ corresponds to the ground state of the model for $\vk L\gg 1$  \eqref{spec-sol0} (see  also \eqref{spec-sol1} below).

However, in both cases, we begin our consideration with special generating functions of a more general form. In the case of the field correlation function,
we consider a matrix element of the form
\be{GF-Fie}
\GFF_N(x;\bv,\bu)=\langle0|\bs{C}(\bv)\; \Psi^\dagger(x)\Psi(0)\;\bs{B}(\bu)|0\rangle.
\ee
Here $\bs{B}(\bu)|0\rangle$ and $\langle0|\bs{C}(\bv)$ respectively are ordinary and twisted on-shell Bethe vectors (that is,  $\bu=\{u_1,\dots,u_N\}$ and $\bv=\{v_1,\dots,v_N\}$ respectively satisfy ordinary and twisted Bethe equations). We introduce the roots of twisted Bethe equations to avoid possible singularities at\footnote{%
In general, the expressions for correlation functions are not singular, but individual terms may have poles at $v_j=u_j$.} $v_j=u_j$.
At the end of the calculations, we set $v_j=u_j+i\beta/L$, $\bu=\bla$, and consider the limit $\beta\to0$.

In the case of the density correlation function, the generating function is more complex. First, we introduce an operator of the number of particles on the interval $[0,x]$
\be{OpNum}
Q_x=\int_0^x j(z)\,\dd z.
\ee
Then it is easy to show that
\be{CF-Den-Qx}
\langle0|\bs{C}(\bla)\; j(x)j(0)\;\bs{B}(\bla)|0\rangle
= \frac12\frac{\partial^2}{\partial x^2}\frac{\partial^2}{\partial \alpha^2}\langle0|\bs{C}(\bla)\;e^{\alpha Q_x}\;\bs{B}(\bla)|0\rangle\Big|_{\alpha=0}.
\ee
Therefore, we start our consideration with a generating function of the form
\be{GF-Den}
\GFD_N(x;\bv,\bu)=\langle0|\bs{C}(\bv)\;e^{\alpha Q_x}\;\bs{B}(\bu)|0\rangle.
\ee
Similarly to equation \eqref{GF-Fie}, here $\bs{B}(\bu)|0\rangle$ and $\langle0|\bs{C}(\bv)$ respectively are ordinary and twisted on-shell Bethe vectors. At the end of the calculations, we make the same limiting procedure as in the case of the field correlation function.

In \cite{KitKMST07} (see also a detailed derivation in \cite{Sla22L}), the following representation for the generating function $\GFD_N(x;\bv,\bu)$ was obtained:
\begin{equation}\label{expQ1-for}
\GFD_N(x;\bv,\bu)=e^{\alpha N}\sum_{\substack{\bv\mapsto\{\bv_{\so},\bv_{\st}\}\\\bu\mapsto\{\bu_{\so},\bu_{\st}\}}}
\frac{r^{(1)}(\bv_{\st})}{r^{(1)}(\bu_{\st})}
f(\bu_{\st},\bu_{\so})f(\bv_{\so},\bv_{\st})\Bomj_{N_{\so}}(\bv_{\so},\bu_{\so}|\beta-\alpha)\Bomj_{N_{\st}}(\bu_{\st},\bv_{\st}|-\alpha).
\end{equation}
Similarly to equation \eqref{SP1}, the sum in \eqref{expQ1-for} is taken over all possible partitions  of the set $\bu$ into two disjoint subsets $\bu_{\so}$ and $\bu_{\st}$ and the set $\bv$ into two disjoint subsets $\bv_{\so}$ and $\bv_{\st}$, so that $\#\bu_{\so}=\#\bv_{\so}=N_{\so}$, $\#\bu_{\st}=\#\bv_{\st}=N_{\st}$, $N_{\so}+N_{\st}=N$, and $N_{\so}=0,\dots,N$.
We also assume that $\Bomj_0(\emptyset,\emptyset|\beta)=1$ by definition. The function $r^{(1)}(z)$ is given by \eqref{rr}.

As we noted above, in the case of general solutions of the ordinary and twisted Bethe equations $\bu$ and $\bv$, the sum over partitions \eqref{expQ1-for} generates a representation in the form of a multiple Cauchy-type contour integral \cite{KitMST05}. However, as we will see below, when calculating the correlation function of densities in the attractive Lieb--Liniger model at zero temperature, most of the terms in this sum turn out to be exponentially small in the $\vk L\gg 1$ regime.

A similar representation exists for the generating function $\GFF_N(x;\bv,\bu)$
\begin{equation}\label{13new-Mk-act3}
\GFF_N(x;\bv,\bu)=-\vk
\sum_{\substack{ \bu\mapsto\{\bu_{\so},\bu_{\st}\} \\ \bv\mapsto\{\bv_{\so},\bv_{\st}\} } }
 \; \frac{r^{(1)}(\bv_{\st})}{r^{(1)}(\bu_{\st})}\;f(\bu_{\st},\bu_{\so})
 f(\bv_{\so},\bv_{\st})
 \Bom_{N_{\so}}(\bv_{\so},\bu_{\so}|\beta)\,  \Bom_{N_{\st}}(\bu_{\st},\bv_{\st}|0).
\end{equation}
The sum in \eqref{13new-Mk-act3} is taken over partitions  of the set $\bu$ into two disjoint subsets $\bu_{\so}$ and $\bu_{\st}$ and the set $\bv$ into two disjoint subsets $\bv_{\so}$ and $\bv_{\st}$, however, restrictions for cardinalities of the subsets are different from ones in \eqref{expQ1-for}.
Now $N_{\so}=\#\bu_{\so}=\#\bv_{\so}+1$, $N_{\st}=\#\bu_{\st}+1=\#\bv_{\st}$, and $N_{\so}+N_{\st}=N$. Thus, $0< N_{\so},\;N_{\st} <N$.
We derive representation \eqref{13new-Mk-act3} in appendix~\ref{A-RGF}.

\section{Special cases of the functions $\Bomj_n(\bv,\bu|\beta)$ and $\Bom_n(\bv,\bu|\beta)$\label{S-SCFG}}

Below we will need a very particular case of the functions $\Bomj_n(\bv,\bu|\beta)$ and $\Bom_n(\bv,\bu|\beta)$. We will find an explicit
expression for these functions in this particular case.

\subsection{Special values of the parameters $\bu$\label{SS-SVPu}}

\begin{prop}\label{P-Bompart} Let $u_{k+1}=u_k+\eta$, for $k=1,\dots, n-1$. Then
\be{Fin-sol}
\Bomj_n(\bv,\bu|\beta)=\left(e^\beta-1\right)n!g(u_1,\bv)g(u_n,\bv)\sum_{k=0}^{n-1}\frac{\binom{n-1}{k}e^{k\beta}(-1)^{n-1-k}}{ g(u_{k+1},\bv)},
\ee
and
\be{Expexp-pc1}
\Bom_n(\bv,\bu|\beta)=n!g(u_1,\bv)g(u_n,\bv)\sum_{k=0}^{n-1}\frac{\binom{n-1}{k}e^{k\beta}(-1)^{n-1-k}}{ g(u_{k+1},\bv)}.
\ee
\end{prop}

\begin{remark}\label{rem-a1}
Formulas \eqref{Fin-sol} and \eqref{Expexp-pc1} look identical except for the $\left(e^\beta-1\right)$ factor in \eqref{Fin-sol}. One should  remember, however, that $\bv=\{v_1,\dots,v_n\}$ for $\Bomj_n(\bv,\bu|\beta)$, but
$\bv=\{v_1,\dots,v_{n-1}\}$ for $\Bom_n(\bv,\bu|\beta)$. Therefore,
\be{vv}
\begin{aligned}
&g(u_k,\bv)=\prod_{j=1}^ng(u_k,v_j),&\qquad &\text{for }\quad\Bomj_n(\bv,\bu|\beta),\\
&g(u_k,\bv)=\prod_{j=1}^{n-1}g(u_k,v_j),&\qquad &\text{for }\quad\Bom_n(\bv,\bu|\beta).
\end{aligned}
\ee
\end{remark}

\begin{remark}\label{rem-a2} Representations similar to \eqref{Fin-sol} and \eqref{Expexp-pc1} were obtained in \cite{CalC07b}. However,  the case $\bu=\bla$ was considered
in that work. Therefore, the condition $u_{k+1}=u_k+\eta$ was satisfied approximately in the $\vk L\gg 1$ regime. The authors of \cite{CalC07b} calculated the leading term of the asymptotics. In this paper, we propose a different method based on recursions \eqref{Fn-rec} and \eqref{rec-G1}. 
\end{remark}

The proofs of formulas \eqref{Fin-sol} and \eqref{Expexp-pc1} are very similar. We will give details of the proof of \eqref{Expexp-pc1}.

\textsl{Proof.} We use induction in $n$. For $n=1$, we have $\bv=\emptyset$, hence, $g(u,\bv)=1$. Then equation \eqref{Expexp-pc1} gives us
$\Bom_1(\emptyset,u|\beta)=1$, which coincides with \eqref{in-cond}.

Assume that \eqref{Expexp-pc1} holds for some $n-1$. Then we can use recursion \eqref{rec-G1} to construct $\Bom_n(\bv,\bu|\beta)$.

It is easy to see that under the condition of the proposition,
\be{2-f}
\begin{aligned}
f(u_j,\bu_j)&=\prod_{\substack{k=1\\k\ne j}}^n\frac{j-k+1}{j-k}=n\delta_{jn},\\
f(\bu_j,u_j)&=\prod_{\substack{k=1\\k\ne j}}^n\frac{j-k-1}{j-k}=n\delta_{j1}.
\end{aligned}
\ee
Then recursion \eqref{rec-G1} takes the form
\begin{multline}\label{Fin-rec}
\Bom_{n}(\bv,\bu|\beta)=n g(u_1,v_{n-1})
e^\beta f(u_1,\bv_{n-1})\Bom_{n-1}(\bv_{n-1},\bu_1|\beta)\\
-ng(u_n,v_{n-1}) f(\bv_{n-1},u_n)\Bom_{n-1}(\bv_{n-1},\bu_n|\beta).
\end{multline}
Substituting expression \eqref{Expexp-pc1} for $\Bom_{n-1}$ into the rhs of the recursion
\eqref{Fin-rec} we find
\begin{multline}\label{Fin-rec1}
\Bom_{n}(\bv,\bu|\beta)=n! g(u_1,v_{n-1})
e^\beta f(u_1,\bv_{n-1})g(u_2,\bv_{n-1})g(u_n,\bv_{n-1})\sum_{k=0}^{n-2}\frac{\binom{n-2}{k}e^{k\beta}(-1)^{n-k}}{ g(u_{k+2},\bv_{n-1})}\\
%
-n!g(u_n,v_{n-1}) f(\bv_{n-1},u_n)g(u_1,\bv_{n-1})g(u_{n-1},\bv_{n-1})\sum_{k=0}^{n-2}\frac{\binom{n-2}{k}e^{k\beta}(-1)^{n-k}}{ g(u_{k+1},\bv_{n-1})}.
\end{multline}
Due to $u_{k+1}=u_k+\eta$, we have
\be{canc}
\begin{aligned}
&g(u_1,v_{n-1})f(u_1,\bv_{n-1})g(u_2,\bv_{n-1})g(u_n,\bv_{n-1})=g(u_1,\bv)g(u_n,\bv_{n-1}),\\
&g(u_n,v_{n-1}) f(\bv_{n-1},u_n)g(u_1,\bv_{n-1})g(u_{n-1},\bv_{n-1})=g(u_1,\bv_{n-1})g(u_n,\bv).
\end{aligned}
\ee
Thus, equation \eqref{Fin-rec1} takes the form
\be{Fin-rec2}
\Bom_{n}(\bv,\bu|\beta)=n!g(u_1,\bv)g(u_n,\bv)\mathcal{A}_n,
\ee
where
\be{An}
\mathcal{A}_n=\frac1{ g(u_n,v_{n-1})}\sum_{k=0}^{n-2}\frac{\binom{n-2}{k}e^{(k+1)\beta}(-1)^{n-k}}{ g(u_{k+2},\bv_{n-1})}
-\frac1{ g(u_1,v_{n-1})}\sum_{k=0}^{n-2}\frac{\binom{n-2}{k}e^{k\beta}(-1)^{n-k}}{ g(u_{k+1},\bv_{n-1})}.
\ee
Shifting $k$ to $k-1$ in the first sum we obtain
\be{An-1}
\mathcal{A}_n=\frac{e^{\beta(n-1)}}{ g(u_n,\bv)}+\frac{(-1)^{n-1}}{ g(u_1,\bv)}
+\sum_{k=1}^{n-2}\frac{e^{k\beta}(-1)^{n-1-k}}{ g(u_{k+1},\bv_{n-1})} \left[\frac{\binom{n-2}{k-1}}{ g(u_n,v_{n-1})}
+\frac{\binom{n-2}{k}}{ g(u_1,v_{n-1})}\right].
\ee
It remains to check that
\begin{multline}\label{check}
\frac{\binom{n-2}{k-1}}{ g(u_n,v_{n-1})}
+\frac{\binom{n-2}{k}}{ g(u_1,v_{n-1})}\\
=\frac1\eta\Big(\binom{n-2}{k-1}(u_n-v_{n-1})
+\binom{n-2}{k} (u_1-v_{n-1})\Big)=\frac{\binom{n-1}{k}}{ g(u_{k+1},v_{n-1})},
\end{multline}
and we arrive at
\be{An-2}
\mathcal{A}_n=\sum_{k=0}^{n-1}\frac{e^{k\beta}\binom{n-1}{k}(-1)^{n-1-k}}{ g(u_{k+1},\bv)} ,
\ee
thus reproducing \eqref{Expexp-pc1}. \qed

\begin{remark}\label{rem-a3} In the case of $\bu=\bla$ and $\vk L\gg 1$, conditions \eqref{2-f} become approximate. They are satisfied up to terms of order $O\big((\vk L)^{-\infty}\big)$. Therefore, recursion \eqref{Fin-rec} is also valid up to terms of order $O\big((\vk L)^{-\infty}\big)$. As a result, the functions $\Bomj_{N}(\bv,\bla|\beta)$ and $\Bom_{N}(\bv,\bla|\beta)$ are respectively specified by formulas \eqref{Fin-sol} and \eqref{Expexp-pc1} up to exponentially small corrections.
\end{remark}

\subsection{Expansion over $\beta$\label{SS-EOB}}

In the representations for the correlation functions, we need either the values of the functions $\Bomj_{n}(\bv,\bu|\beta)$ and $\Bom_{n}(\bv,\bu|\beta)$ at $\beta=0$ or the values of the derivatives of these functions at $\beta=0$. Therefore, we can expand
representations \eqref{Fin-sol} and \eqref{Expexp-pc1} in series with respect to the parameter $\beta$ and then restrict ourselves to only the first few terms of the expansions. Then formulas \eqref{Fin-sol} and \eqref{Expexp-pc1} are simplified even more.

Indeed, it is easy to see that under the conditions of proposition~\ref{P-Bompart},
\be{binom}
g(u_{k+1},\bu_{k+1})=\frac{(-1)^{n-1-k}}{(n-1)!}\binom{n-1}{k}.
\ee
Thus, the sums over $k$ in \eqref{Fin-sol} and \eqref{Expexp-pc1} can be written as follows:
\be{sum-k1}
\sum_{k=0}^{n-1}\frac{\binom{n-1}{k}e^{k\beta}(-1)^{n-1-k}}{ g(u_{k+1},\bv)}=(n-1)!\sum_{k=1}^n\frac{g(u_k,\bu_k)}{g(u_k,\bv)}e^{\beta/g(u_k,u_1)}.
\ee
The sums in the rhs of \eqref{sum-k1} are easily calculated in lower orders in $\beta$. For example, in the case of the function $\Bomj_{n}(\bv,\bu|\beta)$, we have
\be{sum-k2}
\sum_{k=1}^n\frac{g(u_k,\bu_k)}{g(u_k,\bv)}e^{\beta/g(u_k,u_1)}=\frac1\eta\sum_{k=1}^n\frac{\prod_{j=1}^n (u_k-v_j)}{\prod_{\substack{j=1\\ j\ne k}}^n (u_k-u_j)}+O(\beta)
=\frac1{2\pi i\eta}\oint \prod_{j=1}^n\frac{z-v_j}{z-u_j}\,\dd z+O(\beta).
\ee
Here the integration is carried out along an anticlockwise oriented contour surrounding the parameters $\bu$. Evaluating this integral by the residue at infinity we obtain
\be{sum-k3}
\sum_{k=1}^n\frac{g(u_k,\bu_k)}{g(u_k,\bv)}e^{\beta/g(u_k,u_1)}
=\frac1{\eta}\sum_{j=1}^n(u_j-v_j)+O(\beta).
\ee

Similarly, in the case of the function $\Bom_{n}(\bv,\bu|\beta)$, we have
\begin{multline}\label{sum-k4}
\sum_{k=1}^n\frac{g(u_k,\bu_k)}{g(u_k,\bv)}e^{\beta/g(u_k,u_1)}
=\frac1{2\pi i}\oint \frac{\prod_{j=1}^{n-1}(z-v_j)}{\prod_{j=1}^n(z-u_j)}\left(1+\frac\beta\eta(z-u_1)\right)\,\dd z+O(\beta^2)\\
=1+\frac\beta\eta\left(\sum_{j=2}^n u_j-\sum_{j=1}^{n-1} v_j\right)+O(\beta^2).
\end{multline}
Thus,
\be{Beta-expan}
\begin{aligned}
&\Bomj_n(\bv,\bu|\beta)=\frac\beta{\eta}n!(n-1)!g(u_1,\bv)g(u_n,\bv) \sum_{j=1}^n(u_j-v_j)+O(\beta^2),\\
&\Bom_n(\bv,\bu|\beta)=n!(n-1)!g(u_1,\bv)g(u_n,\bv)\left(1+\beta(n-1)+\frac\beta\eta\sum_{j=1}^{n-1}(u_j- v_j)\right)+O(\beta^2).
\end{aligned}
\ee

Using this method, one can find higher orders in the $\beta$-expansion. However, the corresponding coefficients become more cumbersome.

\subsection{Special values of the parameters $\bv$\label{SS-SVPv}}

Let the parameters $\bu$ still satisfy the condition of proposition~\ref{P-Bompart}. We now impose restrictions on the parameters $\bv$.
Namely, let now $v_{k+1}=v_k+\eta$. Here $k=1,\dots,n-1$ for the function $\Bomj_{n}(\bv,\bu|\beta)$, and $k=1,\dots,n-2$ for the
function $\Bom_{n}(\bv,\bu|\beta)$. We also assume that
\be{v1u1}
v_1-u_1=\eta s,
\ee
where $s$ is a complex number. In this case, the functions $\Bomj_{n}(\bv,\bu|\beta)$ and $\Bom_{n}(\bv,\bu|\beta)$ no longer depend on two sets of parameters $\bv$ and $\bu$, but on one complex number $s$. Therefore we denote them by $\Bomj_n(s|\beta)$ and $\Bom_n(s|\beta)$:
\be{hatS}
\begin{aligned}
&\Bomj_n(s|\beta)=\Bomj_n(\{v_1,v_1+\eta\dots,v_1+(n-1)\eta\}, \{u_1,u_1+\eta\dots,u_1+(n-1)\eta\}|\beta),\\
&\Bom_n(s|\beta)=\Bom_n(\{v_1,v_1+\eta\dots,v_1+(n-2)\eta\}, \{u_1,u_1+\eta\dots,u_1+(n-1)\eta\}|\beta).
\end{aligned}
\ee

Using
\be{prodg}
\begin{aligned}
&g(u_{k+1},\bv)=\prod_{j=1}^{n}\frac1{k-j+1-s}=\frac{(-1)^{n}\Gamma(s-k)}{\Gamma(n+s-k)},&\qquad &\text{for}\quad \Bomj_n(s|\beta),\\
&g(u_{k+1},\bv)=\prod_{j=1}^{n-1}\frac1{k-j+1-s}=\frac{(-1)^{n-1}\Gamma(s-k)}{\Gamma(n-1+s-k)},&\qquad &\text{for}\quad \Bom_n(s|\beta),
\end{aligned}
\ee
we obtain
\be{Phis}
\begin{aligned}
&\Bomj_n(s|\beta)=\frac{(-1)^{n} n!\left(e^\beta-1\right)}{s\Gamma(n+s)\Gamma(n-s)}H^{\mathbf{J}}_n,\\
&\Bom_n(s|\beta)=\frac{(-1)^{n-1} n! }{\Gamma(n-1+s)\Gamma(n-s)}H^{\mathbf{\Psi}}_n,
\end{aligned}
\ee
where
\be{Hn-0}
\begin{aligned}
&H^{\mathbf{J}}_n=\sum_{k=0}^{n-1} e^{k\beta}\binom{n-1}{k}\Gamma(k+1-s)\Gamma(n+s-k),\\
&H^{\mathbf{\Psi}}_n=\sum_{k=0}^{n-1}e^{k\beta}\binom{n-1}{k}\Gamma(k+1-s)\Gamma(n-1+s-k).
\end{aligned}
\ee
The sums over $k$ in \eqref{Hn-0} reduce to the hypergeometric function
\be{HGF}
\HGF(a,b;c;z)=\sum_{k=0}^\infty \frac{[a]_k[b]_k}{[c]_k}\frac{z^k}{k!}, \qquad [x]_k=\frac{\Gamma(x+k)}{\Gamma(x)}.
\ee
Actually, below we will need only the first nontrivial term of this series
\be{HGF-1}
\HGF(a,b;c;z)=1+ \frac{ab}{c}z+ O(z^2).
\ee

Let us calculate the sum $H^{\mathbf{J}}_n$ in \eqref{Hn-0}. Assume that $-1< \Re s <1$. Then
\begin{multline}\label{HGFJ}
H^{\mathbf{J}}_n=
n!\sum_{k=0}^{n-1} e^{k\beta}\binom{n-1}{k}\int_0^1 z^{k-s}(1-z)^{n+s-k-1}\,\dd z\\
=n!\int_0^1z^{-s}(1-z)^{s}\left(1+z( e^{\beta}-1)\right)^{n-1}\,\dd z=
\frac{\pi n!}{\sin\pi s}\HGF(1-s,1-n;2;1-e^\beta).
\end{multline}
Here we have used (see \cite{PruBM86})
\be{Prud}
\int_0^1z^{a-1}(1-z)^{b-1}(1+zt)^{c}\,\dd z=
\frac{\Gamma(a)\Gamma(b)}{\Gamma(a+b)}\HGF(a,-c;a+b;-t).
\ee
Since we are dealing with a finite sum over $k$ in equation \eqref{Hn-0}, the result obtained is naturally analytically continued  from the region $-1< \Re s <1$ to the whole complex plane of $s$.

The sum $H^{\mathbf{\Psi}}_n$ is calculated similarly:
\be{HGF0}
H^{\mathbf{\Psi}}_n=(n-1)!\HGF(1-s,1-n;1;1-e^\beta).
\ee
Thus,
\be{Phis-1}
\begin{aligned}
&\Bomj_n(s|\beta)=(-1)^{n}\frac{\pi (n!)^2\left(e^\beta-1\right)}{\sin\pi s\Gamma(n+s)\Gamma(n-s)}\HGF(1-s,1-n;2;1-e^\beta),\\[8pt]
&\Bom_n(s|\beta)=(-1)^{n-1}\frac{\pi n! (n-1)!}{\sin\pi s\Gamma(n-1+s)\Gamma(n-s)}\HGF(1-s,1-n;1;1-e^\beta).
\end{aligned}
\ee

\section{Ground  state correlation functions\label{S-GSCF}}

In this section, we proceed directly to calculating the correlation functions.
We will see that the ground state correlation functions for the Lieb--Liniger model with attraction are described precisely by the special cases of the functions $\Bomj_n(s|\beta)$ and $\Bom_n(s|\beta)$ given in \eqref{Phis-1}.

From now on we consider the $\vk L\gg 1$ regime and set $\bu=\bla=\{\lambda_1,\dots,\lambda_N\}$, where
\be{spec-sol1}
\lambda_j=\lambda_1+\eta(j-1)+\eta\epsilon_j,\qquad  \epsilon_j=O\left((\vk L)^{-\infty}\right), \qquad j=1,\dots,N,
\ee
and $\lambda_1=\eta(1-N)/2$. Recall that $\eta=i\vk$, where $-\vk$ is the coupling constant of the model.
We also set $\bv=\bmu=\{\mu_1,\dots,\mu_N\}$, where
$\mu_j=\lambda_j+i\beta/L$. We consider the limit $\beta\to 0$ at the end of calculations.

\subsection{Norm of the ground state\label{SS-NGS}}

We have mentioned already in the beginning of section~\ref{SS-DFF} that the function $\Bomj_N(\bv,\bu|\beta)$ describes the scalar product of a twisted dual on-shell Bethe vector $\langle0|\bs{C}(\bv)$ and an ordinary on-shell Bethe vector $\bs{B}(\bu)|0\rangle$. Therefore,
\be{PhisNorm0}
\langle0|\bs{C}(\bla)\bs{B}(\bla)|0\rangle=\lim_{\beta\to 0}\Bomj_N(\bmu,\bla|\beta).
\ee
Let
\be{gamma}
\gamma=\frac{i\beta}{\eta L}.
\ee
Then up to exponentially small corrections, $\Bomj_N(\bmu,\bla|\beta)=\Bomj_N(\gamma|\beta)$, and using \eqref{Phis-1} we find
\be{PhisNorm1}
\langle0|\bs{C}(\bla)\bs{B}(\bla)|0\rangle=\lim_{\beta\to 0}\frac{(-1)^{N}\pi (N!)^2\left(e^\beta-1\right)}{\sin\pi \gamma
\Gamma(N+\gamma)\Gamma(N-\gamma)}\HGF(1-\gamma,1-N;2;1-e^\beta)
+O\left((\vk L)^{-\infty}\right).
\ee
Taking the limit we immediately obtain
\be{PhisN2}
\langle0|\bs{C}(\bla)\bs{B}(\bla)|0\rangle=(-1)^{N}\vk L N^2+O\left((\vk L)^{-\infty}\right).
\ee
We recall that the scalar product \eqref{PhisN2} differs from the true squared norm just in the factor $(-1)^{N}$ (see \eqref{Hermit}).

\subsection{Correlation function of fields\label{SS-CFF}}

We now consider the generating function of fields. Setting $\bu=\bla$ and $\bv=\bmu$ in \eqref{13new-Mk-act3} we obtain
\begin{equation}\label{GFF-00}
\GFF_N(x;\bmu,\bla)=-\vk
\sum_{\substack{ \bla\mapsto\{\bla_{\so},\bla_{\st}\} \\ \bmu\mapsto\{\bmu_{\so},\bmu_{\st}\} } }
 \; \frac{r^{(1)}(\bmu_{\st})}{r^{(1)}(\bla_{\st})}\;f(\bla_{\st},\bla_{\so})
 f(\bmu_{\so},\bmu_{\st})
 \Bom_{N_{\so}}(\bmu_{\so},\bla_{\so}|\beta)\,  \Bom_{N_{\st}}(\bla_{\st},\bmu_{\st}|0).
\end{equation}

Consider the product $f(\bla_{\st},\bla_{\so})$ in \eqref{GFF-00}.
\begin{prop}\label{P1}
Let $\lambda_\ell\in\bla_{\st}$. Then $\lambda_{\ell+1}\in\bla_{\st}$, otherwise the corresponding partition gives exponentially small contribution to the
generating function.
\end{prop}

\textsl{Proof.} Suppose that $\lambda_{\ell+1}\in\bla_{\so}$. Then the product $f(\bla_{\st},\bla_{\so})$ contains a factor $f(\lambda_\ell,\lambda_{\ell+1})$. This factor
vanishes up to $O\left((\vk L)^{-\infty}\right)$ terms due to \eqref{spec-sol1}. All other factors in  \eqref{GFF-00} have finite limits at $\lambda_{\ell+1}=\lambda_\ell+\eta$.
Hence, this partition provides an exponentially small contribution. \qed

It follows from this proposition that if $\lambda_{\ell}\in\bla_{\st}$, then all $\lambda_{k}\in\bla_{\st}$ for $k>\ell$.
\begin{cor}
Let $\ell+1$ be the minimal subscript in the set $\bla_{\st}$. Then $\bla_{\st}=\{\lambda_{\ell+1},\dots,\lambda_N\}$
and $\bla_{\so}=\{\lambda_{1},\dots,\lambda_\ell\}$. All other partitions give exponentially small contributions to the
generating function.
\end{cor}

\textsl{Proof.} This statement follows obviously from proposition~\ref{P1}.

Similarly, we can prove that the product $f(\bmu_{\so},\bmu_{\st})$ gives finite contributions only for partitions of the form
$\bmu_{\st}=\{\mu_{1},\dots,\mu_{\ell'}\}$  and $\bmu_{\so}=\{\mu_{\ell'+1},\dots,\mu_{N}\}$, $\ell'=1,\dots,N$.
Since $\#\bmu_{\st}=\#\bla_{\st}+1$ we conclude that $\ell'=N-\ell+1$.

Thus, up to the terms of order $O\left((\vk L)^{-\infty}\right)$, the only partitions giving rise to the result are
\be{partn0}
\begin{aligned}
&\bla_{\so}=\{\lambda_{1},\dots,\lambda_\ell\},&\qquad&\bla_{\st}=\{\lambda_{\ell+1},\dots,\lambda_N\},\\
&\bmu_{\so}=\{\mu_{N-\ell+2},\dots,\mu_{N}\},&\qquad & \bmu_{\st}=\{\mu_{1},\dots,\mu_{N-\ell+1}\},
\end{aligned} \qquad \ell=1,\dots,N.
\ee
We see that the huge sum over the partitions has been reduced to $N$ terms corresponding to the values $\ell=1,\dots,N$.

Then we obtain
\be{ff}
\begin{aligned}
&f(\bla_{\st},\bla_{\so})=\binom{N}{\ell}+O\left((\vk L)^{-\infty}\right), \\
&f(\bmu_{\so},\bmu_{\st})=\binom{N}{\ell-1}+O\left((\vk L)^{-\infty}\right),
\end{aligned}
\ee
and
\be{ll}
\frac{r^{(1)}(\bmu_{\st})}{r^{(1)}(\bla_{\st})}=\exp\Big\{-ix\lambda_1+i\eta x(\ell-1)(N-\ell)-i\eta\gamma x(N-\ell+1)\Big\}
+O\left((\vk L)^{-\infty}\right).
\ee

It is also easy to see that the parameters $\bla$ and $\bmu$ in all subsets are distributed uniformly with an interval $\eta$ up to exponentially small corrections. Therefore, we deal with a particular case of functions $\Bom_{N_{\so}}(\bmu_{\so},\bla_{\so}|\beta)$ and $\Bom_{N_{\st}}(\bla_{\st},\bmu_{\st}|0)$ when
they reduce to the function $\Bom_n(s|\beta)$ (see \eqref{Phis-1}):
\be{FF}
\begin{aligned}
&\Bom_{N_{\so}}(\bmu_{\so},\bla_{\so}|\beta)=\Bom_{\ell}(N+1-\ell+\gamma|\beta)+O\left((\vk L)^{-\infty}\right),\\
&\Bom_{N_{\st}}(\bla_{\st},\bmu_{\st}|0) =\Bom_{N+1-\ell}(\ell-\gamma|0)+O\left((\vk L)^{-\infty}\right).
\end{aligned}
\ee
Thus, equation \eqref{GFF-00} takes the form
\begin{multline}\label{expQ1-for1}
\GFF_N(x;\bmu,\bla)=-\vk\sum_{\ell=1}^N\binom{N}{\ell} \binom{N}{\ell-1}\exp\Big\{-ix\lambda_1+i\eta x(\ell-1)(N-\ell)-i\eta\gamma x(N-\ell+1)\Big\}\\
\times\Bom_{\ell}(N+1-\ell+\gamma|\beta)\Bom_{N+1-\ell}(\ell-\gamma|0)+O\left((\vk L)^{-\infty}\right).
\end{multline}
Substituting here \eqref{Phis-1} for the functions $\Bom$ we after simple algebra arrive at
\begin{equation}\label{expQ1-for1a1}
\GFF_N(x;\bmu,\bla)=\frac{(-1)^{N-1}\vk\pi (N!)^2e^{-ix\lambda_1}}{\sin\pi\gamma\Gamma(N-\gamma)\Gamma(N+\gamma)}
\sum_{\ell=1}^Ne^{i\eta x(\ell-1)(N-\ell)}\Lambda^{\mathbf{\Psi}}_\ell+O\left((\vk L)^{-\infty}\right),
\end{equation}
where
\begin{equation}\label{LamPsi}
\Lambda^{\mathbf{\Psi}}_\ell=(2\ell-N-1-\gamma)e^{-i\eta\gamma x(N-\ell+1)}
\HGF(\ell-N-\gamma,1-\ell;1;1-e^\beta).
\end{equation}

The terms of this sum are singular\footnote{The only exception is the term for $\ell=(N+1)/2$ that occurs for $N$ odd.} in the limit $\beta\to0$.
However, it is easy to see that the combination $\Lambda^{\mathbf{\Psi}}_\ell+\Lambda^{\mathbf{\Psi}}_{N-\ell+1}$
has a well-defined limit $\beta\to0$. Thus, we replace  $\ell$ with $N-\ell+1$ in \eqref{expQ1-for1a1} and add the resulting
sum to the original one. We obtain
\be{corf-F1}
\GFF_N(x;\bmu,\bla)=\frac{(-1)^{N-1}\vk\pi (N!)^2e^{-ix\lambda_1}}{2\sin\pi\gamma\Gamma(N-\gamma)\Gamma(N+\gamma)}
\sum_{\ell=1}^N  e^{i\eta x(\ell-1)(N-\ell)}  M^{\mathbf{\Psi}}_\ell(\beta)+O\left((\vk L)^{-\infty}\right),
\ee
where
\begin{multline}\label{Mell}
M^{\mathbf{\Psi}}_\ell(\beta)=\Lambda^{\mathbf{\Psi}}_\ell+\Lambda^{\mathbf{\Psi}}_{N-\ell+1}=(2\ell-N-1-\gamma)e^{-i\eta\gamma x(N-\ell+1)}
 \HGF(\ell-N-\gamma,1-\ell;1;1-e^\beta)\\
 -(2\ell-N-1+\gamma)e^{-i\eta\gamma x\ell}
 \HGF(1-\ell-\gamma,\ell-N;1;1-e^\beta).
\end{multline}
We see that $M^{\mathbf{\Psi}}_\ell(0)=0$. Hence, taking the $\beta\to 0$ limit (that is, sending $\bmu\to\bla$) we arrive at
\be{corf-F2}
\GFF_N(x;\bla,\bla)=(-1)^{N-1}\vk^2 L N^2e^{-\vk x(N-1)/2}
\sum_{\ell=1}^N  e^{-\vk x(\ell-1)(N-\ell)}  \frac12
\frac{\dd}{\dd\beta} M^{\mathbf{\Psi}}_\ell(\beta)\Bigr|_{\beta=0}+O\left((\vk L)^{-\infty}\right).
\ee
Using \eqref{HGF-1} we find
\be{exp-HGF}
\begin{aligned}
&\HGF(\ell-N-\gamma,1-\ell;1;1-e^\beta)=1-\beta(\ell-1)(N-\ell)+O(\beta^2),\\
&\HGF(1-\ell-\gamma,\ell-N;1;1-e^\beta)=1-\beta(\ell-1)(N-\ell)+O(\beta^2).
\end{aligned}
\ee
Hence,
\be{Mb-1}
M^{\mathbf{\Psi}}_\ell(\beta)=i\eta\gamma x(N-\ell+1)^2-2\gamma+O(\beta^2),
 \ee
 and
 \be{Mb-2}
 \frac{\dd}{\dd\beta} M^{\mathbf{\Psi}}_\ell(\beta)\Bigr|_{\beta=0}=-\frac{2}{\vk L}-\frac xL (N-2\ell+1)^2.
 \ee
Then equation \eqref{corf-F2} yields
\begin{multline}\label{corf-F3}
\GFF_N(x;\bla,\bla)=(-1)^{N}\vk N^2e^{-\vk x(N-1)/2}\\
\times\sum_{\ell=1}^N  e^{-\vk x(\ell-1)(N-\ell)}
\left(1+\frac{\vk x}2(N-2\ell+1)^2\right)+O\left((\vk L)^{-\infty}\right).
\end{multline}
It remains to divide this result by \eqref{PhisN2}, and we arrive at representation \eqref{CF-Fie-res1}.

\subsection{Correlation function of densities\label{SS-CFD}}

The correlation function of densities is calculated similarly. There are a few minor differences that we will note below.

First of all, we rewrite representation \eqref{expQ1-for}, putting $\bu=\bla$ and $\bv=\bmu$ in it
\begin{equation}\label{GFD-00}
\GFD_N(x;\bmu,\bla)=e^{\alpha N}\sum_{\substack{\bmu\mapsto\{\bmu_{\so},\bmu_{\st}\}\\\bla\mapsto\{\bla_{\so},\bla_{\st}\}}}
\frac{r^{(1)}(\bmu_{\st})}{r^{(1)}(\bla_{\st})}
f(\bla_{\st},\bla_{\so})f(\bmu_{\so},\bmu_{\st})\Bomj_{N_{\so}}(\bmu_{\so},\bla_{\so}|\beta-\alpha)\Bomj_{N_{\st}}(\bla_{\st},\bmu_{\st}|-\alpha).
\end{equation}
An analysis completely similar to that carried out in the previous section shows that finite contributions to the generating function $\GFD_N(x;\bmu,\bla)$ are given only by partitions of the following form:
\be{partn1}
\begin{aligned}
&\bla_{\so}=\{\lambda_{1},\dots,\lambda_\ell\},&\qquad&\bla_{\st}=\{\lambda_{\ell+1},\dots,\lambda_N\},\\
&\bmu_{\so}=\{\mu_{N-\ell+1},\dots,\mu_{N}\},&\qquad & \bmu_{\st}=\{\mu_{1},\dots,\mu_{N-\ell}\},
\end{aligned} \qquad \ell=0,\dots,N.
\ee
The slight difference with \eqref{partn1} is explained by the fact that now $\#\bla_{\so}=\#\bmu_{\so}$ and
$\#\bla_{\st}=\#\bmu_{\st}$.

Then we obtain
\be{ffJ}
\begin{aligned}
&f(\bla_{\st},\bla_{\so})=\binom{N}{\ell}+O\left((\vk L)^{-\infty}\right), \\
&f(\bmu_{\so},\bmu_{\st})=\binom{N}{\ell}+O\left((\vk L)^{-\infty}\right),
\end{aligned}
\ee
and
\be{llJ}
\frac{r^{(1)}(\bmu_{\st})}{r^{(1)}(\bla_{\st})}=\exp\Big\{i\eta x(\ell-\gamma)(N-\ell)\Big\}
+O\left((\vk L)^{-\infty}\right).
\ee
We also have
\be{FFJ}
\begin{aligned}
&\Bomj_{N_{\so}}(\bmu_{\so},\bla_{\so}|\beta-\alpha)=\Bomj_{\ell}(N-\ell+\gamma|\beta-\alpha)+O\left((\vk L)^{-\infty}\right),&\qquad& \ell>0,\\
&\Bomj_{N_{\st}}(\bla_{\st},\bmu_{\st}|-\alpha) =\Bomj_{N-\ell}(\ell-\gamma|-\alpha)+O\left((\vk L)^{-\infty}\right),&\qquad& \ell<N.
\end{aligned}
\ee
Recall also the $\Bomj_0(\emptyset,\emptyset|\beta)=1$ by definition.

Let us consider contributions to \eqref{GFD-00} provided by the partitions with $\bla_{\so}=\bmu_{\so}=\emptyset$
\be{LJ0}
\Lambda^{\mathbf{J}}_0=e^{\alpha N}\frac{r^{(1)}(\bmu)}{r^{(1)}(\bla)}\Bomj_{N}(\bla,\bmu|-\alpha)
=e^{\alpha N+x\gamma\vk N}\Bomj_{N}(-\gamma|-\alpha)+O\left((\vk L)^{-\infty}\right),
\ee
and the partitions with $\bla_{\st}=\bmu_{\st}=\emptyset$
\be{LJN}
\Lambda^{\mathbf{J}}_N=e^{\alpha N}\Bomj_{N}(\bmu,\bla|\beta-\alpha)=e^{\alpha N}\Bomj_{N}(\gamma|\beta-\alpha)+O\left((\vk L)^{-\infty}\right).
\ee
We have due to \eqref{Phis-1}
\be{L0NJ}
\begin{aligned}
&\Lambda^{\mathbf{J}}_0=(-1)^{N-1}\frac{\pi (N!)^2\left(e^{-\alpha}-1\right)e^{\alpha N+x\gamma\vk N}}
{\sin\pi \gamma\Gamma(N+\gamma)\Gamma(N-\gamma)}\HGF(1+\gamma,1-N;2;1-e^{-\alpha}),\\[8pt]
&\Lambda^{\mathbf{J}}_N=(-1)^{N}\frac{\pi (N!)^2\left(e^{\beta-\alpha}-1\right)e^{\alpha N}}
{\sin\pi \gamma\Gamma(N+\gamma)\Gamma(N-\gamma)}\HGF(1-\gamma,1-N;2;1-e^{\beta-\alpha}).
\end{aligned}
\ee
Obviously, the sum $\Lambda^{\mathbf{J}}_0+\Lambda^{\mathbf{J}}_N$ has a well defined $\beta\to0$ limit. Moreover, it is easy to see that
\be{comb0N}
\lim_{\beta\to 0}\Big(\Lambda^{\mathbf{J}}_0+\Lambda^{\mathbf{J}}_N\Big)=ax+b,
\ee
where $a$ and $b$ are $x$-independent. Thus, this contribution will be killed by the second $x$-derivative (see \eqref{CF-Den-Qx}).

The remaining partitions give the following contribution to equation \eqref{GFD-00}:
\begin{equation}\label{expQ1-for1a1J}
\GFD_N(x;\bmu,\bla)=\frac{(-1)^{N}\pi (N!)^2\left(e^{\beta-\alpha}-1\right)\left(e^{-\alpha}-1\right)e^{\alpha N}}{\sin\pi\gamma\Gamma(N-\gamma)\Gamma(N+\gamma)}
\sum_{\ell=1}^{N-1}e^{i\eta x\ell(N-\ell)}\Lambda^{\mathbf{J}}_\ell+O\left((\vk L)^{-\infty}\right),
\end{equation}
where
\begin{equation}\label{LamJ}
\Lambda^{\mathbf{J}}_\ell=(N-2\ell+\gamma)e^{-i\eta\gamma x(N-\ell)}
\HGF(\ell+1-N-\gamma,1-\ell;2;1-e^{\beta-\alpha})\HGF(1-\ell+\gamma,\ell+1-N;2;1-e^{-\alpha}).
\end{equation}
We see that again $\Lambda^{\mathbf{J}}_\ell+\Lambda^{\mathbf{J}}_{N-\ell}$ is non-singular at $\beta\to 0$. Thus, similarly to the case considered above, we obtain
\be{corf-F1J}
\GFD_N(x;\bmu,\bla)=\frac{(-1)^{N}\pi (N!)^2\left(e^{\beta-\alpha}-1\right)\left(e^{-\alpha}-1\right)e^{\alpha N}}{2\sin\pi\gamma\Gamma(N-\gamma)\Gamma(N+\gamma)}
\sum_{\ell=1}^{N-1} e^{i\eta x\ell(N-\ell)}  M^{\mathbf{J}}_\ell(\beta,\alpha)+O\left((\vk L)^{-\infty}\right),
\ee
where $M^{\mathbf{J}}_\ell(\beta,\alpha)=\Lambda^{\mathbf{J}}_\ell+\Lambda^{\mathbf{J}}_{N-\ell}$.
%
%
 %
 %
%
It is easy to see that $M^{\mathbf{J}}_\ell(0,\alpha)=0$. Hence, taking the $\beta\to 0$ limit  we arrive at
\be{corf-F2J}
\GFD_N(x;\bla,\bla)=(-1)^{N}\vk L N^2\left(e^{-\alpha}-1\right)^2 e^{\alpha N}
\sum_{\ell=1}^N  e^{-\vk x\ell(N-\ell)}  \frac12
\frac{\dd}{\dd\beta} M^{\mathbf{J}}_\ell(\beta,\alpha)\Bigr|_{\beta=0}+O\left((\vk L)^{-\infty}\right).
\ee

To simplify further formulas, we calculate the second $\alpha$-derivative of the generating function at $\alpha=0$ (see \eqref{CF-Den-Qx}). It is clear that this derivative acts only on the factor $\left(e^{-\alpha}-1\right)^2$. In the rest of equation \eqref{corf-F2J}, we simply set $\alpha=0$:
\be{corf-F3J}
\frac12\frac{\partial^2}{\partial\alpha^2}\GFD_N(x;\bla,\bla)\Bigr|_{\alpha=0}=(-1)^{N}\vk L N^2
\sum_{\ell=1}^N  e^{-\vk x\ell(N-\ell)}  \frac12
\frac{\dd}{\dd\beta} M^{\mathbf{J}}_\ell(\beta,0)\Bigr|_{\beta=0}+O\left((\vk L)^{-\infty}\right),
\ee
where
\begin{multline}\label{MellJ0}
M^{\mathbf{J}}_\ell(\beta,0)=(N-2\ell+\gamma)e^{x\vk\gamma(N-\ell)}
 \HGF(\ell+1-N-\gamma,1-\ell;2;1-e^{\beta})\\
 -(N-2\ell-\gamma)e^{x\vk\gamma\ell}
 \HGF(1-\ell-\gamma,\ell+1-N;2;1-e^{\beta}).
\end{multline}
Using \eqref{HGF-1} we obtain
\be{Mb-1J}
M^{\mathbf{J}}_\ell(\beta,0)=2\gamma+\vk x\gamma(N-2\ell)^2+O(\beta^2),
 \ee
 and hence,
 \be{Mb-2J}
 \frac12\frac{\dd}{\dd\beta} M^{\mathbf{J}}_\ell(\beta,0)\Bigr|_{\beta=0}=\frac{1}{\vk L}\left(1+\frac{\vk x}2 (N-2\ell)^2\right).
 \ee
Then equation \eqref{corf-F2J} yields
\be{corf-F4J}
\frac12\frac{\partial^2}{\partial\alpha^2}\GFD_N(x;\bla,\bla)\Bigr|_{\alpha=0}=(-1)^{N} N^2\sum_{\ell=1}^N  e^{-\vk x\ell(N-\ell)}
\left(1+\frac{\vk x}2(N-2\ell)^2\right)+O\left((\vk L)^{-\infty}\right).
\ee
Normalizing this result by \eqref{PhisN2} and acting with the second $x$-derivative we arrive at representation \eqref{veryfin-res}.

\section*{Conclusion}

We studied the ground state correlation functions of densities and fields in the model of one-dimensional bosons with attraction in the $\vk L\gg 1$ regime. The key formulas in our approach are representations \eqref{expQ1-for} and \eqref{13new-Mk-act3}. They define generating functions for correlations of densities and fields. These representations are obtained using the ABA method. They are valid for any coupling constant, as well as for any eigenstate of the Hamiltonian.

In the case of the Lieb--Liniger model with repulsion, equations \eqref{expQ1-for} and \eqref{13new-Mk-act3} ultimately lead to representations for the correlation functions in the form of multiple integrals \cite{KitKMST07}. The analysis of these integrals is a very difficult task. To date, based on these integral representations, it has been possible to obtain only the asymptotic behavior of correlation functions at a large distance $x$ \cite{KitKMST09}.

However, for the attractive Lieb--Liniger model in the $\vk L\gg 1$ regime, equations \eqref{expQ1-for} and \eqref{13new-Mk-act3} are dramatically simplified. Of the entire sum by partitioning of $2N$ roots of the Bethe equations, only $N$ terms make a finite contribution to the correlation. Moreover, these terms are also simplified, since all the determinants included in them are explicitly calculated. As a result, we obtain very simple and explicit results for the correlation functions.

The reason for the simplifications noted above is the very special form of solution of the Bethe equations corresponding to the ground state. The roots of this solution form a so-called string, and the interval between the roots is equal to $i\vk$ up to exponentially small corrections. It is thanks to this structure of the ground state that it is possible to explicitly take into account all contributions to the correlation functions.

Since representations \eqref{expQ1-for}, \eqref{13new-Mk-act3} are valid for any eigenstate of the Hamiltonian, we can also use them to calculate temperature correlation functions. However, we have to deal with states that contain a mixture of strings of different lengths in this case. Therefore, we no longer have such significant simplifications, despite the fact that string solutions of the Bethe equations are still determined up to exponentially small corrections in the $\vk L\gg 1$ regime. Many more terms survive in the sums over partitions, and these terms themselves have a more complex form. This observation further highlights the fact that it is the remarkable ground state structure that is responsible for the dramatic simplifications of the correlation functions at zero temperature in the attractive Lieb--Lineger model. Apparently, this is the main difference between the model under consideration and $1+1$-dimensional models of quantum field theory, in which significant simplifications are also observed in the large system size limit.

Unfortunately, the method used in this paper is not suitable for computing dynamical correlation functions. Either it should be significantly modified, or resort to the form factor decomposition, as was done in \cite{CalC07a,CalC07b}. In the latter case, the question arises about summation over the complete set of intermediate states. Recall that in \cite{CalC07a,CalC07b}, contributions from only two types of excitations were taken into account. It is quite possible that taking into account all types of excitations when calculating dynamic correlations will lead to simpler results.

\section*{Acknowledgements}
I am grateful to A.~Pogrebkov and V.~Spiridonov for fruitful discussions.
This work was performed at the Steklov International Mathematical Center and supported by the Ministry of Science and Higher Education of the Russian Federation (agreement no. 075-15-2022-265).

\appendix

\section{Representation for the generating function\label{A-RGF}}

In this appendix, we study matrix element \eqref{GF-Fie}. The first task is to calculate the actions of the fields $\Psi^\dagger(x)$ and $\Psi(0)$ on (dual) Bethe vectors in the form of linear combinations of new (dual) Bethe vectors. Thus, we express the matrix element \eqref{GF-Fie} in terms of a linear combination of scalar products.

\subsection{Representation for the matrix element in terms of the scalar products \label{AA-RMESP}}

First, we rewrite \eqref{Psiact} as follows:
\be{Psiact-1}
\Psi(0)\bs{B}(\bu)|0\rangle=\sqrt{\vk}\sum_{\bu\mapsto\{\bu_{0},\bu_{\sth}\}}r(\bu_0)f(\bu_{\sth},\bu_0)\bs{B}(\bu_{\sth})|0\rangle.
\ee
Here the sum is taken over partitions of the set $\bu$ into subsets $\bu_{0}$ and $\bu_{\sth}$ so that $\#\bu_{0}=1$. Using Bethe equations \eqref{BE}
we obtain
\be{Psiact-2}
\Psi(0)\bs{B}(\bu)|0\rangle=\sqrt{\vk}\sum_{\bu\mapsto\{\bu_{0},\bu_{\sth}\}}f(\bu_0,\bu_{\sth})\bs{B}(\bu_{\sth})|0\rangle.
\ee
Using \eqref{13-IK} we find
\be{Psiact-3}
\Psi(0)\bs{B}(\bu)|0\rangle=\sqrt{\vk}\sum_{\bu\mapsto\{\bu_{0},\bu_{\so},\bu_{\st}\}}f(\bu_0,\bu_{\so})f(\bu_0,\bu_{\st})
f(\bu_{\st},\bu_{\so})
r^{(2)}(\bu_{\so})\bs{B}^{(1)}(\bu_{\so})|0\rangle^{(1)}\bs{B}^{(2)}(\bu_{\st})|0\rangle^{(2)}.
\ee
Here we have an additional partition of the subset $\bu_{\sth}$ into $\bu_{\so}$ and $\bu_{\st}$.

Finally, taking into account \eqref{rr} and using again Bethe equations we arrive at
\be{Psiact-4}
\Psi(0)\bs{B}(\bu)|0\rangle=\sum_{\bu\mapsto\{\bu_{0},\bu_{\so},\bu_{\st}\}}
\frac{\sqrt{\vk}}{r^{(1)}(\bu_{\so})}
f(\bu_{\so},\bu_{\st}) f(\bu_{\so},\bu_{0})f(\bu_0,\bu_{\st}) \bs{B}^{(1)}(\bu_{\so})|0\rangle^{(1)}\bs{B}^{(2)}(\bu_{\st})|0\rangle^{(2)}.
\ee
Thus, the action of $\Psi(0)$ on the on-shell Bethe vector $\bs{B}(\bu)|0\rangle$ is given by the sum over partitions of the original set $\bu$ into three
subsets $\{\bu_{0},\bu_{\so},\bu_{\st}\}$ with the condition $\#\bu_{0}=1$.

To obtain the action of $\Psi^\dagger(x)$ on the dual twisted on-shell Bethe vector $\bs{C}(\bv)$ we first apply \eqref{13-IK-d}  and
then  use \eqref{Psiact2} in the form
\be{Psiactd-1}
\langle0|\bs{C}^{(2)}(\bv_{\sth})\Psi^\dagger(x)=-\sqrt{\vk}\sum_{\bv_{\sth}\mapsto\{\bv_{0},\bv_{\st}\}}^N f(\bv_0,\bv_{\st})\langle0|\bs{C}^{(2)}(\bv_{\st}).
\ee
Here the sum is taken over partitions of the set $\bv_{\sth}$ into subsets $\bv_{0}$ and $\bv_{\st}$ so that $\#\bv_{0}=1$. Then we obtain
\begin{multline}\label{Psiactd-2}
\langle0| \bs{C}(\bv)\Psi^\dagger(x)=-\sqrt{\vk}\sum_{\bv\mapsto\{\bv_{0},\bv_{\so},\bv_{\st}\}}
r^{(1)}(\bv_{0})r^{(1)}(\bv_{\st})
\; f(\bv_{\so},\bv_{\st}) f(\bv_{\so},\bv_{0}) f(\bv_0,\bv_{\st})\\
\times \langle0|^{(1)}\bs{C}^{(1)}(\bv_{\so})\langle0|^{(2)}\bs{C}^{(2)}(\bv_{\st}).
\end{multline}
Combining \eqref{Psiact-4} and \eqref{Psiactd-2}  we arrive at
\begin{multline}\label{1-repr-GF}
\GFF_N(x;\bv,\bu)=-\vk\sum_{\substack{\bu\mapsto\{\bu_{0},\bu_{\so},\bu_{\st}\}\\ \bv\mapsto\{\bv_{0},\bv_{\so},\bv_{\st}\}}}
f(\bu_{\so},\bu_{\st}) f(\bu_{\so},\bu_{0})f(\bu_0,\bu_{\st})
\; f(\bv_{\so},\bv_{\st})f(\bv_{\so},\bv_{0}) f(\bv_0,\bv_{\st})\\
\times \frac{r^{(1)}(\bv_{0})r^{(1)}(\bv_{\st})}{r^{(1)}(\bu_{\so})}
\langle0|^{(1)}\bs{C}^{(1)}(\bv_{\so}) \bs{B}^{(1)}(\bu_{\so})|0\rangle^{(1)}
\langle0|^{(2)}\bs{C}^{(2)}(\bv_{\st}) \bs{B}^{(2)}(\bu_{\st})|0\rangle^{(2)}.
\end{multline}
The sum is taken over partitions $\bu\mapsto\{\bu_0,\bu_{\so},\bu_{\st}\}$ and  $\bv\mapsto\{\bv_0,\bv_{\so},\bv_{\st}\}$ such that $\#\bu_{\so}=\#\bv_{\so}=N_{\so}$,
$\#\bu_{\st}=\#\bv_{\st}=N_{\st}$ (otherwise, the scalar products in \eqref{1-repr-GF}  vanish), and $\#\bu_{0}=\#\bv_{0}=1$.

\subsection{Representation for the matrix element in terms of the field form factors \label{AA-RMEFFF}}

From now on, the calculations almost completely repeat those that were done to calculate the generating function $\GFD_N$ \eqref{GF-Den} in \cite{KitKMST07,Sla22L}.
We should now calculate the  scalar products of the partial Bethe vectors:
\begin{equation}\label{13new-SP-k}
\bs{S}_m^{(j)}(\bv|\bu)= \langle0|^{(j)}\bs{C}^{(j)}(\bv) \bs{B}^{(j)}(\bu)|0\rangle^{(j)},\qquad j=1,2,\qquad m=N_{\so},N_{\st}.
\end{equation}
For their calculation, we can use the general formula \eqref{SP1}
replacing there the functions $r(z)$ with the functions $r^{(1)}(z)$ in the case of the
scalar product $\bs{S}_{N_ {\so}}^{(1)}$, and $r(z)$ with $r^{(2)}(z)$ for the scalar product $\bs{S}_{N_ {\st}}^{(2)}$. We will have new partitions
into new subsets. Because of the large number of such subsets, we will reject the use of Roman numerals and mark
the subsets with bold superscripts. Then we obtain for the scalar product $\bs{S}_{N_ {\so}}^{(1)}$
\begin{equation}\label{13new-SP1}
 \bs{S}_{N_{\so}}^{(1)}(\bv_{\so}|\bu_{\so})=
 \sum_{\substack{\bv_{\so}\mapsto\{\bv^{\bs{1}},\bv^{\bs{3}}\}\\\bu_{\so}\mapsto\{\bu^{\bs{1}},\bu^{\bs{3}}\}}}
r^{(1)}(\bv^{\bs{3}})r^{(1)}(\bu^{\bs{1}})
\mathbb{K}_{N_{1}}( \bv^{\bs{1}}|\bu^{\bs{1}})\mathbb{K}_{N_{3}}( \bu^{\bs{3}}|\bv^{\bs{3}})\,f(\bu^{\bs{3}},\bu^{\bs{1}})f(\bv^{\bs{1}},\bv^{\bs{3}}).
\end{equation}
The sum in \eqref{13new-SP1} is taken over all possible partitions $\bu_{\so}\mapsto\{\bu^{\bs{1}},\bu^{\bs{3}}\}$ and $\bv_{\so}\mapsto\{\bv^{\bs{1}},\bv^{\bs{3}}\}$,
such that $\#\bu^{\bs{1}}=\#\bv^{\bs{1}}=N_{1}$, $\#\bu^{\bs{3}}=\#\bv^{\bs{3}}=N_{3}$, $N_{1}+N_{3}=N_{\so}$,  and $N_{1}=1,\dots,N_{\so}$.

Similarly,
\begin{equation}\label{13new-SP2}
 \bs{S}_{N_{\st}}^{(2)}(\bu_{\st}|\bv_{\st})=
 \sum_{\substack{\bv_{\st}\mapsto\{\bv^{\bs{2}},\bv^{\bs{4}}\}\\
\bu_{\st}\mapsto\{\bu^{\bs{2}},\bu^{\bs{4}}\}}}
r^{(2)}(\bv^{\bs{4}})r^{(2)}(\bu^{\bs{2}})
\mathbb{K}_{N_{2}}( \bv^{\bs{2}}|\bu^{\bs{2}})\mathbb{K}_{N_{4}}( \bu^{\bs{4}}|\bv^{\bs{4}})
f(\bu^{\bs{4}},\bu^{\bs{2}})f(\bv^{\bs{2}},\bv^{\bs{4}}).
\end{equation}
Here the notation is similar to the one in \eqref{13new-SP1}.

Now we express the functions $r^{(2)}(z)$ in terms of
the functions $r^{(1)}(z)$ and $r(z)$ via \eqref{rr} and use (twisted) Bethe equations:
\begin{equation}\label{13new-BE-v2}
r^{(2)}(\bu^{\bs{2}})=\frac{r(\bu^{\bs{2}})}{r^{(1)}(\bu^{\bs{2}})}=\frac{1}{r^{(1)}(\bu^{\bs{2}})}\;
\frac{f(\bu^{\bs{2}},\bu^{\bs{0}})f(\bu^{\bs{2}},\bu^{\bs{1}})f(\bu^{\bs{2}},\bu^{\bs{3}})f(\bu^{\bs{2}},\bu^{\bs{4}})}
{f(\bu^{\bs{0}},\bu^{\bs{2}})f(\bu^{\bs{1}},\bu^{\bs{2}})f(\bu^{\bs{3}},\bu^{\bs{2}})f(\bu^{\bs{4}},\bu^{\bs{2}})}.
\end{equation}
and
\begin{equation}\label{13new-BE-u4}
r^{(2)}(\bv^{\bs{4}})=\frac{r(\bv^{\bs{4}})}{r^{(1)}(\bv^{\bs{4}})}=\frac{e^{N_4\beta}}{r^{(1)}(\bv^{\bs{4}})}\;
\frac{f(\bv^{\bs{4}},\bv^{\bs{0}})f(\bv^{\bs{4}},\bv^{\bs{1}})f(\bv^{\bs{4}},\bv^{\bs{2}})f(\bv^{\bs{4}},\bv^{\bs{3}})}
{f(\bv^{\bs{0}},\bv^{\bs{4}})f(\bv^{\bs{1}},\bv^{\bs{4}})f(\bv^{\bs{2}},\bv^{\bs{4}})f(\bv^{\bs{3}},\bv^{\bs{4}})},
\end{equation}
where $\bu^{\bs{0}}=\bu_0$  and $\bv^{\bs{0}}=\bv_0$. Then the scalar product $ \bs{S}_{N_{\st}}^{(2)}(\bv_{\st}|\bu_{\st})$ takes the form
\begin{multline}\label{13new-SP2-mod}
 \bs{S}_{N_{\st}}^{(2)}(\bu_{\st}|\bv_{\st})=
 \sum_{\substack{\bv_{\st}\mapsto\{\bv^{\bs{2}},\bv^{\bs{4}}\}\\
\bu_{\st}\mapsto\{\bu^{\bs{2}},\bu^{\bs{4}}\}}}
\frac{e^{N_4\beta}}{r^{(1)}(\bv^{\bs{4}})r^{(1)}(\bu^{\bs{2}})}
\,\mathbb{K}_{N_{2}}( \bv^{\bs{2}}|\bu^{\bs{2}})\mathbb{K}_{N_{4}}( \bu^{\bs{4}}|\bv^{\bs{4}})\\
\times \frac{f(\bv^{\bs{4}},\bv^{\bs{0}})f(\bv^{\bs{4}},\bv^{\bs{1}})f(\bv^{\bs{4}},\bv^{\bs{2}})f(\bv^{\bs{4}},\bv^{\bs{3}})
f(\bu^{\bs{2}},\bu^{\bs{0}})f(\bu^{\bs{2}},\bu^{\bs{1}})f(\bu^{\bs{2}},\bu^{\bs{3}})f(\bu^{\bs{2}},\bu^{\bs{4}})}
{f(\bv^{\bs{0}},\bv^{\bs{4}})f(\bv^{\bs{1}},\bv^{\bs{4}})f(\bv^{\bs{3}},\bv^{\bs{4}})
f(\bu^{\bs{0}},\bu^{\bs{2}})f(\bu^{\bs{1}},\bu^{\bs{2}})f(\bu^{\bs{3}},\bu^{\bs{2}})}.
\end{multline}

We also have
\begin{multline}\label{13new-fact}
\frac{r^{(1)}(\bv_{0})r^{(1)}(\bv_{\st})}{r^{(1)}(\bu_{\so})}f(\bu_{\so},\bu_{\st}) f(\bu_{\so},\bu_{0})f(\bu_{0},\bu_{\st})
\; f(\bv_{\so},\bv_{\st})f(\bv_{\so},\bv_{0}) f(\bv_0,\bv_{\st})
=\frac{r^{(1)}(\bv^{\bs{0}})r^{(1)}(\bv^{\bs{2}})r^{(1)}(\bv^{\bs{4}})}{r^{(1)}(\bu^{\bs{1}})r^{(1)}(\bu^{\bs{3}})} \\
\times f(\bu^{\bs{1}},\bu^{\bs{2}}) f(\bu^{\bs{1}},\bu^{\bs{4}})f(\bu^{\bs{3}},\bu^{\bs{2}})f(\bu^{\bs{3}},\bu^{\bs{4}})
f(\bu^{\bs{1}},\bu^{\bs{0}})f(\bu^{\bs{3}},\bu^{\bs{0}})f(\bu^{\bs{0}},\bu^{\bs{2}})f(\bu^{\bs{0}},\bu^{\bs{4}})\\
\times f(\bv^{\bs{1}},\bv^{\bs{2}})f(\bv^{\bs{1}},\bv^{\bs{4}})f(\bv^{\bs{3}},\bv^{\bs{2}})f(\bv^{\bs{3}},\bv^{\bs{4}})
f(\bv^{\bs{1}},\bv^{\bs{0}})f(\bv^{\bs{3}},\bv^{\bs{0}})f(\bv^{\bs{0}},\bv^{\bs{2}})f(\bv^{\bs{0}},\bv^{\bs{4}}).
\end{multline}

Now we need to substitute formulas \eqref{13new-SP1}, \eqref{13new-SP2-mod}, and \eqref{13new-fact} into  representation \eqref{1-repr-GF}.
We obtain
\begin{multline}\label{13new-Mk-act2}
\GFF_N(x;\bv,\bu)=-\vk\sum
 \;e^{N_4\beta}\frac{r^{(1)}(\bv^{\bs{0}})r^{(1)}(\bv^{\bs{2}})r^{(1)}(\bv^{\bs{3}})}{r^{(1)}(\bu^{\bs{2}})r^{(1)}(\bu^{\bs{3}})}
 \;F_u\, F_v\,\\
 \times
\mathbb{K}_{N_{1}}( \bv^{\bs{1}}|\bu^{\bs{1}})\mathbb{K}_{N_{3}}( \bu^{\bs{3}}|\bv^{\bs{3}})\mathbb{K}_{N_{2}}( \bv^{\bs{2}}|\bu^{\bs{2}})
\mathbb{K}_{N_{4}}( \bu^{\bs{4}}|\bv^{\bs{4}}),
\end{multline}
where
\begin{multline}\label{13new-Fuvla}
F_u=f(\bu^{\bs{3}},\bu^{\bs{1}})
f(\bu^{\bs{2}},\bu^{\bs{0}})f(\bu^{\bs{2}},\bu^{\bs{1}})f(\bu^{\bs{2}},\bu^{\bs{3}}) f(\bu^{\bs{2}},\bu^{\bs{4}})\\
\times
 f(\bu^{\bs{1}},\bu^{\bs{4}})f(\bu^{\bs{3}},\bu^{\bs{4}})
f(\bu^{\bs{1}},\bu^{\bs{0}})f(\bu^{\bs{3}},\bu^{\bs{0}})f(\bu^{\bs{0}},\bu^{\bs{4}}),
\end{multline}
and
\begin{multline}\label{13new-Fuvmu}
F_v=f(\bv^{\bs{1}},\bv^{\bs{3}})
f(\bv^{\bs{4}},\bv^{\bs{0}})f(\bv^{\bs{4}},\bv^{\bs{1}})f(\bv^{\bs{4}},\bv^{\bs{2}})f(\bv^{\bs{4}},\bv^{\bs{3}})\\
\times f(\bv^{\bs{1}},\bv^{\bs{2}})f(\bv^{\bs{3}},\bv^{\bs{2}})
f(\bv^{\bs{1}},\bv^{\bs{0}})f(\bv^{\bs{3}},\bv^{\bs{0}})f(\bv^{\bs{0}},\bv^{\bs{2}}).
\end{multline}
The sum in \eqref{13new-Mk-act2}  is taken over partitions of the original sets $\bv$ and $\bu$ into five subsets each:
\begin{equation}\label{13new-part}
\bu\mapsto\{\bu^{\bs{0}},\bu^{\bs{1}},\bu^{\bs{2}},\bu^{\bs{3}},\bu^{\bs{4}}\},\qquad
\bv\mapsto\{\bv^{\bs{0}},\bv^{\bs{1}},\bv^{\bs{2}},\bv^{\bs{3}},\bv^{\bs{4}}\}.
\end{equation}
The restrictions are imposed only on the cardinalities of the subsets, namely, $\#\bv^{\bs{j}}=\#\bu^{\bs{j}}=N_{j}$, $j=0,1,2,3,4$, and $N_0=1$.

Now we introduce new subsets:
\begin{equation}\label{13new-new-part}
\begin{aligned}
&\bu^{\rm i}=\{\bu^{\bs{0}},\bu^{\bs{1}},\bu^{\bs{4}}\},\qquad & \bu^{\rm ii}=\{\bu^{\bs{2}},\bu^{\bs{3}}\},\\
&\bv^{\rm i}=\{\bv^{\bs{1}},\bv^{\bs{4}}\},\qquad & \bv^{\rm ii}=\{\bv^{\bs{0}},\bv^{\bs{2}},\bv^{\bs{3}}\} .
\end{aligned}
\end{equation}
Then, it is easy to see that
\begin{equation}\label{13new-Fuv-2}
\begin{aligned}
&F_u=f(\bu^{\rm ii},\bu^{\rm i})
f(\bu^{\bs{2}},\bu^{\bs{3}})f(\bu^{\bs{1}},\bu^{\bs{4}})
f(\bu^{\bs{1}},\bu^{\bs{0}})f(\bu^{\bs{0}},\bu^{\bs{4}}),\\
&F_v=f(\bv^{\rm i},\bv^{\rm ii})
f(\bv^{\bs{4}},\bv^{\bs{1}})f(\bv^{\bs{3}},\bv^{\bs{2}})
f(\bv^{\bs{3}},\bv^{\bs{0}})f(\bv^{\bs{0}},\bv^{\bs{2}}).
\end{aligned}
\end{equation}
As a result, we obtain
\begin{equation}\label{13new-Mk-act3aa}
\GFF_N(x;\bv,\bu)=-\vk
\sum_{\substack{ \bu\mapsto\{\bu^{\rm i},\bu^{\rm ii}\} \\ \bv\mapsto\{\bv^{\rm i},\bv^{\rm ii}\} } }
 \; \frac{r^{(1)}(\bv^{\rm ii})}{r^{(1)}(\bu^{\rm ii})}\;f(\bu^{\rm ii},\bu^{\rm i})
 f(\bv^{\rm i},\bv^{\rm ii})
 \\
 \Bom_{N_{\rm i}}(\bv^{\rm i},\bu^{\rm i}|\beta)\,  \Bom_{N_{\rm ii}}(\bu^{\rm ii},\bv^{\rm ii}|0),
\end{equation}
 where $N_{\rm i}=N_{\bs{1}}+N_{\bs{4}}+1$ and $N_{\rm ii}=N_{\bs{2}}+N_{\bs{3}}+1$. The functions $\Bom_{N_{\rm i}}(\bv^{\rm i},\bu^{\rm i}|\beta)$ and
 $\Bom_{N_{\rm ii}}(\bu^{\rm ii},\bv^{\rm ii}|0)$ are given as sums over partitions. Namely,
\be{13new-G1}
\Bom_{N_{\rm i}}(\bv^{\rm i},\bu^{\rm i}|\beta)  =\sum_{\substack{ \bu^{\rm i}\mapsto\{\bu^{\bs{0}},\bu^{\bs{1}},\bu^{\bs{4}}\} \\
\bv^{\rm i}\mapsto\{\bv^{\bs{1}},\bv^{\bs{4}}\} } }
 \mathbb{K}_{N_{\bs{1}}}( \bv^{\bs{1}}|\bu^{\bs{1}})\mathbb{K}_{N_{\bs{4}}}( \bu^{\bs{4}}|\bv^{\bs{4}})
e^{N_{\bs{4}}\beta}f(\bv^{\bs{4}},\bv^{\bs{1}})f(\bu^{\bs{1}},\bu^{\bs{4}})f(\bu^{\bs{1}},\bu^{\bs{0}})f(\bu^{\bs{0}},\bu^{\bs{4}}),
 \ee
and
\be{13new-G2}
 \Bom_{N_{\rm ii}}(\bu^{\rm ii},\bv^{\rm ii}|0)=
 \sum_{\substack{ \bu^{\rm ii}\mapsto\{\bu^{\bs{2}},\bu^{\bs{3}}\} \\ \bv^{\rm ii}\mapsto\{\bv^{\bs{0}},\bv^{\bs{2}},\bv^{\bs{3}}\} } }
 \mathbb{K}_{N_{\bs{3}}}( \bu^{\bs{3}}|\bv^{\bs{3}})\mathbb{K}_{N_{\bs{2}}}( \bv^{\bs{2}}|\bu^{\bs{2}})
f(\bu^{\bs{2}},\bu^{\bs{3}})f(\bv^{\bs{3}},\bv^{\bs{2}})f(\bv^{\bs{3}},\bv^{\bs{0}})f(\bv^{\bs{0}},\bv^{\bs{2}}).
\ee

It is easy to see that representation \eqref{13new-Mk-act3aa} coincides (up to the labels of the subsets) with \eqref{GFF-00}. Thus, it remains to prove that the sums over partitions \eqref{13new-G1} and \eqref{13new-G2} do give us  the field form factors and, therefore, they have determinant representation \eqref{detrepG2-a00}, \eqref{tM-a00}.

\subsection{Determinant representation for the field form factor\label{AA-DRFFF}}

Summation over partitions in \eqref{13new-G1} and \eqref{13new-G2} is based on two auxiliary lemmas. These lemmas were proved in \cite{BelPRS12} (see also \cite{Sla22L}) and represent identities for the highest coefficient $\mathbb{K}_N(\bv|\bu)$ \eqref{HC}.

\begin{lemma}\label{main-ident}
Let $\bar\xi$, $\bar y$ and $\bar z$ be sets of complex parameters with $\#\bar y=m_1$,
$\#\bar z=m_2$, and $\#\bar \xi=m_1+m_2$. Then
\begin{equation}\label{Iden-K}
  \sum_{\bar\xi\mapsto\{\bar\xi_{\so},\bar\xi_{\st}\} }
 \mathbb{K}_{m_1}(\bar\xi_{\so}|\bar y)\mathbb{K}_{m_2}(\bar z|\bar\xi_{\st})f(\bar\xi_{\st},\bar\xi_{\so})
 = (-1)^{m_1}f(\bar\xi,\bar y) \mathbb{K}_{m_1+m_2}(\bar y-\eta,\bar z|\bar\xi).
 \end{equation}
The sum is taken with respect to all partitions of the set $\bar\xi$ into disjoint
subsets $\bar\xi_{\so}$ and $\bar\xi_{\st}$ with $\#\bar\xi_{\so}=m_1$ and $\#\bar\xi_{\st}=m_2$.
\end{lemma}

\begin{lemma}\label{Long-Det}
Let $\bar w$ and $\bar\xi$ be two sets of generic complex numbers with $\#\bar w=\#\bar\xi=m$. Let
also $C_1(w)$ and $C_2(w)$ be two arbitrary functions of a complex variable $w$. Then
\begin{multline}\label{SumDet1}
\sum_{\bar w\mapsto\{\bar w_{\so},\bar w_{\st}\} }
\mathbb{K}_m(\bar w_{\so}-\eta, \bar w_{\st}|\bar \xi)f(\bar \xi, \bar w_{\so})f(\bar w_{\st},\bar w_{\so})
C_1(\bar w_{\so})C_2(\bar w_{\st})\\
=\Delta'(\bar\xi)\Delta(\bar w)
\det_m\Bigl(C_2(w_k)t(w_k,\xi_j)h(w_k,\bar\xi)+(-1)^m C_1(w_k)t(\xi_j,w_k)h(\bar\xi,w_k)\Bigr).
\end{multline}
Here the sum is taken over all possible partitions of the set $\bar w$ into disjoint subsets $\bar w_{\so}$
and $\bar w_{\st}$. We have also extended the convention on the  shorthand notation \eqref{SH-prod0} to the products of functions $C_1$ and $C_2$.
\end{lemma}
The proofs of lemma~\ref{main-ident} and lemma~\ref{Long-Det} are given in \cite{BelPRS12} (see also \cite{Sla22L}).

\begin{cor}\label{Long-DetC}
Under the conditions and with the notation of lemma~\ref{Long-Det}
\begin{multline}\label{SumDet2}
\sum_{\bar w\mapsto\{\bar w_{\so},\bar w_{\st}\} }
\mathbb{K}_m(\bar w_{\so}-\eta, \bar w_{\st}|\bar \xi)f(\bar \xi, \bar w_{\so})f(\bar w_{\so},\bar w_{\st})
C_1(\bar w_{\so})C_2(\bar w_{\st})\\
=\Delta'(\bar\xi)\Delta(\bar w)
\det_m\Bigl(C_2(w_k)t(w_k,\xi_j)h(w_k,\bar\xi)+(-1)^m C_1(w_k)\frac{f(w_k,\bar w_k)}{f(\bar w_k, w_k)} t(\xi_j,w_k)h(\bar\xi,w_k)\Bigr).
\end{multline}
\end{cor}

\textsl{Proof.} Let us introduce a function $\tilde C_1(w_k)$ by
\be{C1C1}
\tilde C_1(w_k)= C_1(w_k) \frac{f(w_k,\bar w_k)}{f(\bar w_k, w_k)}.
\ee
Then obviously,
\be{CCC}
C_1(\bar w_{\so}) f(\bar w_{\so},\bar w_{\st})=\tilde C_1(\bar w_{\so}) f(\bar w_{\st},\bar w_{\so}),
\ee
and we can use lemma~\ref{Long-Det}, replacing there $C_1$ with $\tilde C_1$. \qed

Now summation over partitions in \eqref{13new-G1} and \eqref{13new-G2} becomes straightforward. First of all, we note that equations \eqref{13new-G1} and \eqref{13new-G2} differ from each other only in the replacement $\bv\leftrightarrow\bu$ and the subset indices. Therefore, it is enough to calculate one of these sums.

Let  $\bu=\{u_1,\dots,u_n\}$ and $\bv=\{v_1,\dots,v_{n-1}\}$. Define a function $\Bom_n(\bv,\bu|\beta)$ by
\be{Bom-def}
\Bom_n(\bv,\bu|\beta)  =\sum_{\substack{ \bu\mapsto\{\bub{0},\bub{1},\bub{2}\} \\ \bv\mapsto\{\bvb{1},\bvb{2}\} } }
 \mathbb{K}_{n_{\bs{1}}}( \bvb{1}|\bub{1})\mathbb{K}_{n_{\bs{2}}}( \bub{2}|\bvb{2})
e^{n_{\bs{2}}\beta}f(\bvb{2},\bvb{1})f(\bub{1},\bub{2})f(\bub{1},\bub{0})f(\bub{0},\bub{2}).
 \ee
Here the sum is taken over partitions of the set $\bu$ into three subsets $\{\bub{0},\bub{1},\bub{2}\}$ and the set $\bv$ into two
subsets $\{\bvb{1},\bvb{2}\}$. These partitions are independent except that $\#\bub{0}=1$, $ \#\bvb{1}=\#\bub{1}=n_{\bs{1}}$, $ \#\bvb{2}=\#\bub{2}=n_{\bs{2}}=n-n_{\bs{1}}-1$, and $n_{\bs{1}}=0,1,\dots,n-1$.

Obviously, sum \eqref{Bom-def} differs from sum \eqref{13new-G1} only in notation.
Applying lemma~\ref{main-ident} to the sum over partitions of $\bv$ in \eqref{Bom-def} we obtain
\be{G1-der-1}
 \Bom_{n}(\bv,\bu|\beta)=
 \sum_{\bu\mapsto\{\bub{0},\bub{1},\bub{2}\}}
 \mathbb{K}_{n-1}( \{\bub{1}-\eta,\bub{2}\}|\bv)
(-1)^{n_{\bs{1}}}e^{n_{\bs{2}}\beta}f(\bv,\bub{1})f(\bub{1},\bub{2})f(\bub{1},\bub{0})f(\bub{0},\bub{2}).
 \ee
Let us introduce a union $\bub{3}=\{\bub{1}, \bub{2}\}=\bu\setminus \{\bub{0}\}$. Then we can think that the summation in \eqref{G1-der-1} occurs in two stages:
\be{2stage}
\sum_{\bu\mapsto\{\bub{0},\bub{1},\bub{2}\}}=\sum_{\bu\mapsto\{\bub{0},\bub{3}\}}\sum_{\bub{3}\mapsto\{\bub{1},\bub{2}\}}\;.
\ee
In the first stage, we split the set $\bu$ into subsets $\bub{0}$ and $\bub{3}$, so that $\#\bub{0}=1$. In the second stage, the summation occurs over all partitions of the set  $\bub{3}$ into $\bub{1}$ and $\bub{2}$. It is to this sum that we can apply corollary~\ref{Long-DetC}. We set
\be{spec}
C_1(\ub{3}_k)=-f(\ub{3}_k,\bub{0}),\qquad C_2(\ub{3}_k)=e^\beta f(\bub{0},\ub{3}_k),
\ee
where $\ub{3}_k$ means the $k$th element of the subset $\bub{3}$. Then after simple algebra, we obtain
\begin{multline}\label{sumu1u2}
  \sum_{\bub{3}\mapsto\{\bub{1},\bub{2}\}}
 \mathbb{K}_{n-1}( \{\bub{1}-\eta,\bub{2}\}|\bv)
(-1)^{n_{\bs{1}}}e^{n_{\bs{2}}\beta}f(\bv,\bub{1})f(\bub{1},\bub{2})f(\bub{1},\bub{0})f(\bub{0},\bub{2})\\
=\Delta'(\bv)\Delta(\bub{3})h(\bub{3},\bv)f(\bub{0},\bub{3})\det_{n-1}\mathcal{M}_{jk}(\bv,\bub{3}),
 \end{multline}
where $\mathcal{M}_{jk}(\bv,\bub{3})$ is an $(n-1) \times (n-1)$ matrix
\be{tM-M}
\mathcal{M}_{jk}(\bv,\bub{3})=e^\beta t(\ub{3}_k,v_j)- t(v_j,\ub{3}_k)
\frac{h(\ub{3}_k,\bu)h(\bv,\ub{3}_k)}{h(\bu,\ub{3}_k)h(\ub{3}_k,\bv)}.
\ee
Thus, the sum over partitions \eqref{G1-der-1} turns into
\be{G1-der-1a}
 \Bom_{n}(\bv,\bu|\beta)=
\sum_{\bu\mapsto\{\bub{0},\bub{3}\}}
 \Delta'(\bv)\Delta(\bub{3})h(\bub{3},\bv)f(\bub{0},\bub{3})\det_{n-1}\mathcal{M}_{jk}(\bv,\bub{3}),
 \ee
where $\#\bub{0}=1$ and $\bub{3}=\bu\setminus\bub{0}$.

It is easy to see that the remaining sum \eqref{G1-der-1a} is nothing but an expansion of $\det_n\mathcal{M}^{\mathbf{\Psi}}_{jk}$  \eqref{tM-a00} with respect to the last row.
Indeed, let $\bub{0}=u_m$, where $m=1,\dots,n$. Then $\bub{3}=\bu_m$. In its turn, the determinant of the matrix $\mathcal{M}_{jk}$ is a minor of the matrix $\mathcal{M}^{\mathbf{\Psi}}_{jk}$, which is obtained by removing the last row and the $m$th column:
\be{minor}
\det_{n-1}\mathcal{M}_{jk}(\bv,\bub{3})=\det_{\substack{j\ne n\\ k\ne m}}\mathcal{M}^{\mathbf{\Psi}}_{jk}(\bv,\bu).
\ee
Using $f(\bub{0},\bub{3})=g(\bub{0},\bub{3})h(\bub{0},\bu)$ we also have
\be{fgDel}
f(\bub{0},\bub{3})\Delta(\bub{3})=g(u_m,\bu_m)h(u_m,\bu)\Delta(\bu_m)= (-1)^{m+n}h(u_m,\bu)\Delta(\bu).
\ee
Substituting this into \eqref{G1-der-1a} we arrive at
\begin{multline}\label{G1-der-2a}
\sum_{\bu\mapsto\{\bub{0},\bub{3}\}}
 \Delta'(\bv)\Delta(\bub{3})h(\bub{3},\bv)f(\bub{0},\bub{3})\det_{n-1}\mathcal{M}_{jk}(\bv,\bub{3})\\
 =\Delta'(\bv)\Delta(\bu)h(\bu,\bv)\sum_{m=1}^n (-1)^{m+n}\frac{h(u_m,\bu)}{h(u_m,\bv)}\det_{\substack{j\ne n\\ k\ne m}}\mathcal{M}^{\mathbf{\Psi}}_{jk}(\bv,\bu)\\
 =\Delta'(\bv)\Delta(\bu)h(\bu,\bv)\det_{n}\mathcal{M}^{\mathbf{\Psi}}_{jk}(\bv,\bu).
 \end{multline}
Thus,
\be{G1-fin-res}
 \Bom_{n}(\bv,\bu|\beta) =\Delta'(\bv)\Delta(\bu)h(\bu,\bv)\det_{n}\mathcal{M}^{\mathbf{\Psi}}_{jk}(\bv,\bu),
 \ee
and we have reproduced representation \eqref{detrepG2-a00}.

\end{document}